\documentclass[final,onefignum,onetabnum]{siamart171218}

\usepackage{braket,amsfonts}

\usepackage{array}

\usepackage{nicefrac}

\usepackage[caption=false]{subfig}
\usepackage{pgfplots}
\usepackage{ wasysym }

\newsiamthm{claim}{Claim}
\newsiamremark{remark}{Remark}
\newsiamremark{hypothesis}{Hypothesis}
\crefname{hypothesis}{Hypothesis}{Hypotheses}

\usepackage{algorithmic}

\usepackage{graphicx,epstopdf}
\usepackage{wrapfig}

\Crefname{ALC@unique}{Line}{Lines}

\usepackage{amsopn}

\usepackage{amsmath,amssymb}
\usepackage{xspace}
\usepackage{bold-extra}
\usepackage[most]{tcolorbox}

\colorlet{texcscolor}{blue!50!black}
\colorlet{texemcolor}{red!70!black}
\colorlet{texpreamble}{red!70!black}
\colorlet{codebackground}{black!25!white!25}

\lstdefinestyle{siamlatex}{%
  style=tcblatex,
  texcsstyle=*\color{texcscolor},
  texcsstyle=[2]\color{texemcolor},
  keywordstyle=[2]\color{texemcolor},
  moretexcs={cref,Cref,maketitle,mathcal,text,headers,email,url},
}

\tcbset{%
  colframe=black!75!white!75,
  coltitle=white,
  colback=codebackground, 
  colbacklower=white, 
  fonttitle=\bfseries,
  arc=0pt,outer arc=0pt,
  top=1pt,bottom=1pt,left=1mm,right=1mm,middle=1mm,boxsep=1mm,
  leftrule=0.3mm,rightrule=0.3mm,toprule=0.3mm,bottomrule=0.3mm,
  listing options={style=siamlatex}
}

\newtcblisting[use counter=example]{example}[2][]{%
  title={Example~\thetcbcounter: #2},#1}

\newtcbinputlisting[use counter=example]{\examplefile}[3][]{%
  title={Example~\thetcbcounter: #2},listing file={#3},#1}

\DeclareTotalTCBox{\code}{ v O{} }
{ 
  fontupper=\ttfamily\color{black},
  nobeforeafter,
  tcbox raise base,
  colback=codebackground,colframe=white,
  top=0pt,bottom=0pt,left=0mm,right=0mm,
  leftrule=0pt,rightrule=0pt,toprule=0mm,bottomrule=0mm,
  boxsep=0.5mm,
  #2}{#1}

\patchcmd\newpage{\vfil}{}{}{}
\begin{tcbverbatimwrite}{tmp_\jobname_header.tex}
\title{Metaplex networks: influence of the exo-endo structure of complex systems on diffusion\thanks{Submitted to the editors. 
\funding{
G.~E.~R.~was supported by The Maxwell Institute Graduate School in Analysis and its
Applications, a Centre for Doctoral Training funded by the UK Engineering and Physical
Sciences Research Council (grant EP/L016508/01), the Scottish Funding Council, Heriot-Watt
University and the University of Edinburgh.}}}

\author{Ernesto Estrada${}$\thanks{Institute of Applied Mathematics (IUMA), Universidad de Zaragoza, Pedro Cerbuna 12, E-50009 Zaragoza, Spain; ARAID Foundation, Government of Arag\'{o}n, E-50018 Zaragoza, Spain; and Instituto de Ciencias Matematicas e de Computacao, Universidade de Sao Paulo, Caixa Postal 668, 13560-970 Sao Carlos, Sao Paulo, Brazil (\email{estrada66@unizar.es}).}\and 
  Gissell Estrada-Rodriguez\thanks{Maxwell Institute for Mathematical Sciences and Department of Mathematics, Heriot-–Watt University, Edinburgh, EH14 4AS, United Kingdom
    (\email{ge5@hw.ac.uk},
    \email{h.gimperlein@hw.ac.uk}.)}
  \and
  Heiko Gimperlein${}^\ddag$
}

\headers{Networks with internal structure}{G. Estrada-Rodriguez, E. Estrada, and H. Gimperlein}
\end{tcbverbatimwrite}
\title{Metaplex networks: influence of the exo-endo structure of complex systems on diffusion\thanks{Submitted to the editors. 
\funding{
G.~E.~R.~was supported by The Maxwell Institute Graduate School in Analysis and its
Applications, a Centre for Doctoral Training funded by the UK Engineering and Physical
Sciences Research Council (grant EP/L016508/01), the Scottish Funding Council, Heriot-Watt
University and the University of Edinburgh.}}}

\author{Ernesto Estrada${}$\thanks{Institute of Applied Mathematics (IUMA), Universidad de Zaragoza, Pedro Cerbuna 12, E-50009 Zaragoza, Spain; ARAID Foundation, Government of Arag\'{o}n, E-50018 Zaragoza, Spain; and Instituto de Ciencias Matematicas e de Computacao, Universidade de Sao Paulo, Caixa Postal 668, 13560-970 Sao Carlos, Sao Paulo, Brazil (\email{estrada66@unizar.es}).}\and
  Gissell Estrada-Rodriguez\thanks{Maxwell Institute for Mathematical Sciences and Department of Mathematics, Heriot-–Watt University, Edinburgh, EH14 4AS, United Kingdom
    (\email{ge5@hw.ac.uk},
    \email{h.gimperlein@hw.ac.uk}.)}
  \and
  Heiko Gimperlein${}^\dag$
}

\headers{Networks with internal structure}{G. Estrada-Rodriguez, E. Estrada, and H. Gimperlein}

\ifpdf
\hypersetup{ pdftitle={Guide to Using  SIAM'S \LaTeX\ Style} }
\fi

\begin{document}
\maketitle

\begin{tcbverbatimwrite}{tmp_\jobname_abstract.tex}
\begin{abstract}
\noindent In a complex system the interplay between the internal structure of its entities and their interconnection may play a fundamental role in the global functioning of the system. Here, we define the concept of metaplex, which describes such trade-off between internal structure of entities and their interconnections. We then define a dynamical system on a metaplex and study diffusive processes on them. We provide analytical and computational evidences about the role played by the size of the nodes, the location of the internal coupling areas, and the strength and range of the coupling between the nodes on the global dynamics of metaplexes. Finally, we extend our analysis to two real-world metaplexes: a landscape and a brain metaplex. We corroborate that the internal structure of the nodes in a metaplex may dominate the global dynamics (brain metaplex) or play a regulatory role (landscape metaplex) to the influence of the interconnection between nodes. 

\end{abstract}


\end{tcbverbatimwrite}
\begin{abstract}
\noindent In a complex system the interplay between the internal structure of its entities and their interconnection may play a fundamental role in the global functioning of the system. Here, we define the concept of metaplex, which describes such trade-off between internal structure of entities and their interconnections. We then define a dynamical system on a metaplex and study diffusive processes on them. We provide analytical and computational evidences about the role played by the size of the nodes, the location of the internal coupling areas, and the strength and range of the coupling between the nodes on the global dynamics of metaplexes. Finally, we extend our analysis to two real-world metaplexes: a landscape and a brain metaplex. We corroborate that the internal structure of the nodes in a metaplex may dominate the global dynamics (brain metaplex) or play a regulatory role (landscape metaplex) to the influence of the interconnection between nodes.

\end{abstract}




\section{Introduction}\label{sec: introduction}
A complex system is characterized by the existence of many components
which are interdependent on each other \cite{Complex_networks_1,Complex_networks_2,Newman_review}. Each of these components is at the same time characterized
by certain structure, dynamics and function \cite{networks_dynamics},
which influences the global behaviour of the system. The interconnection
between these components represents the exo-skeleton of the complex
system, and it is well characterized by the use of complex networks
\cite{Complex_networks_1,Complex_networks_2,Newman_review}. The internal structure of the corresponding
entities~--~inside the 
nodes of the complex network~--~represents the endo structure
of the system and it is not necessarily a network in itself. Let us
consider some examples, from the vast variety that exists in
nature and society, to illustrate the point. 

\begin{figure}
\begin{centering}
\subfloat[\label{fig: cells}]{\includegraphics[width=0.45\textwidth]{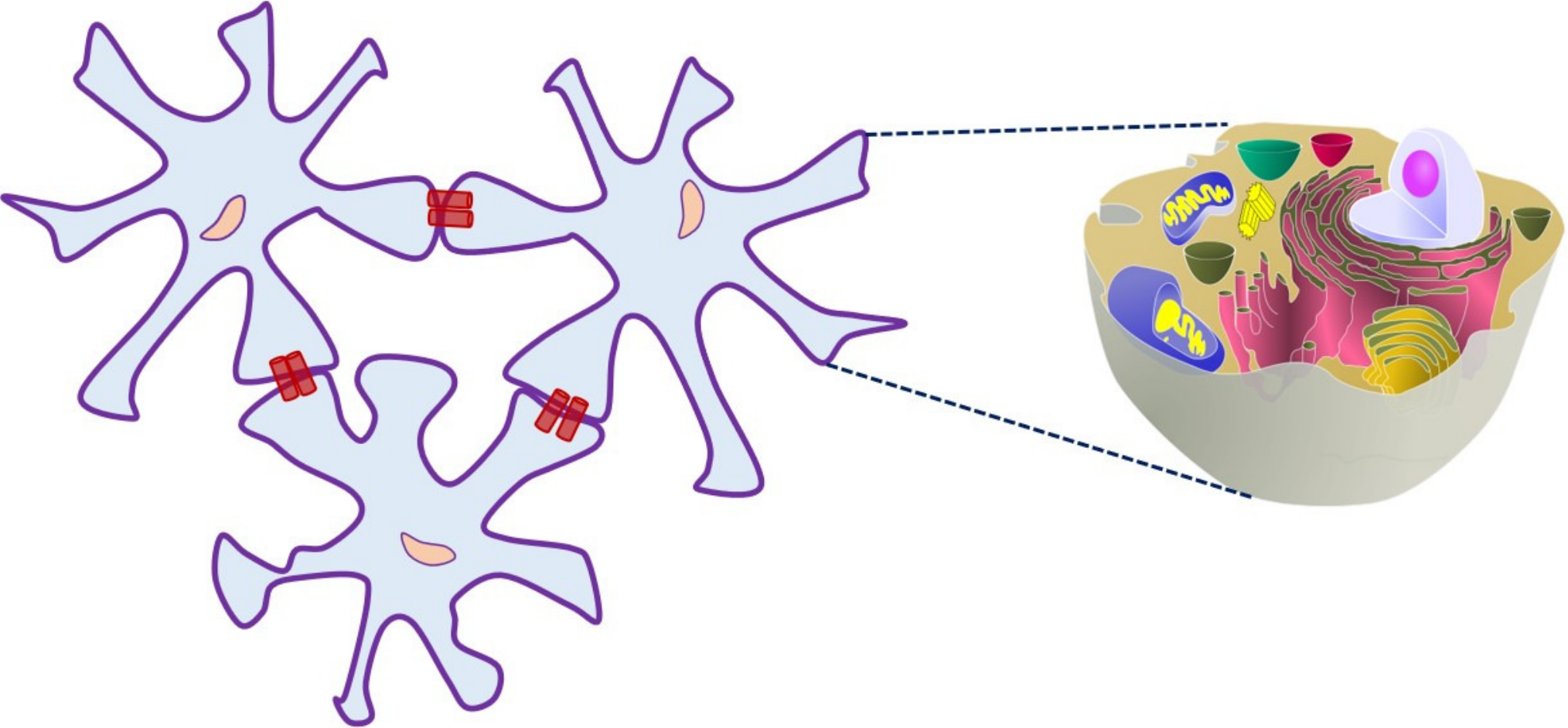}

}\subfloat[\label{fig: circles}]{\includegraphics[width=0.45\textwidth]{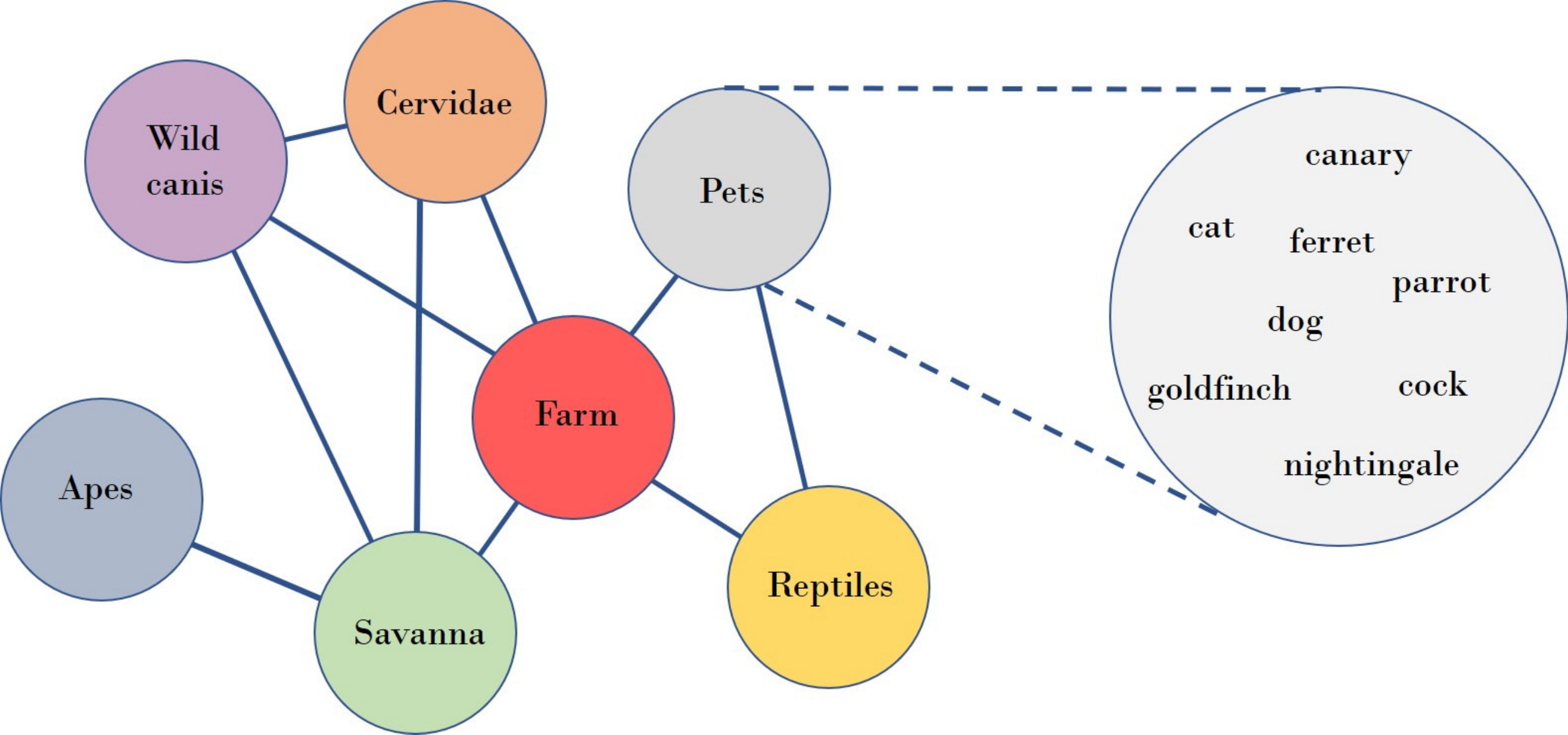}

}
\par\end{centering}
\begin{centering}
\subfloat[\label{fig: elephants}]{\includegraphics[width=0.45\textwidth]{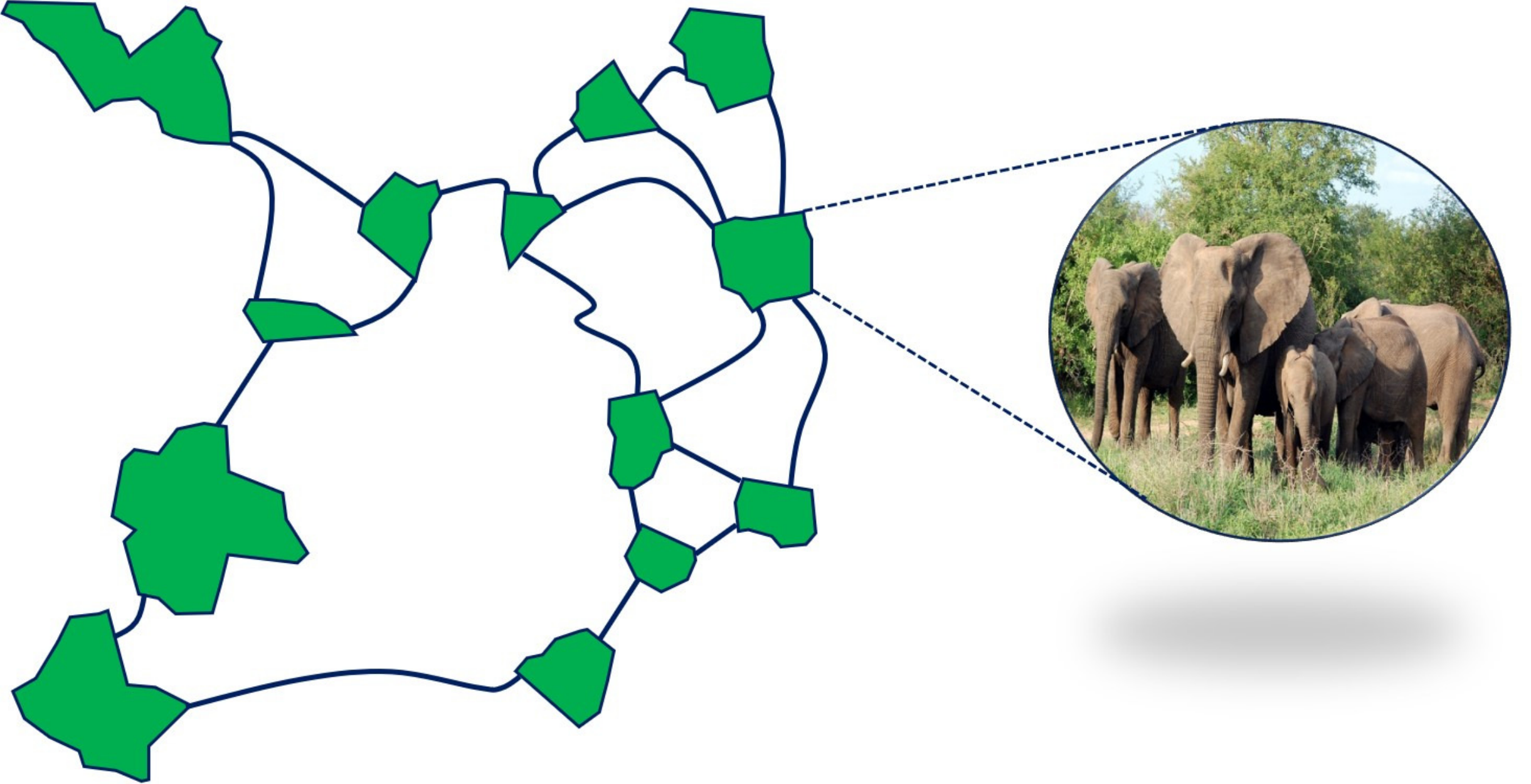}

}\subfloat[\label{fig: cell}]{\includegraphics[width=0.45\textwidth]{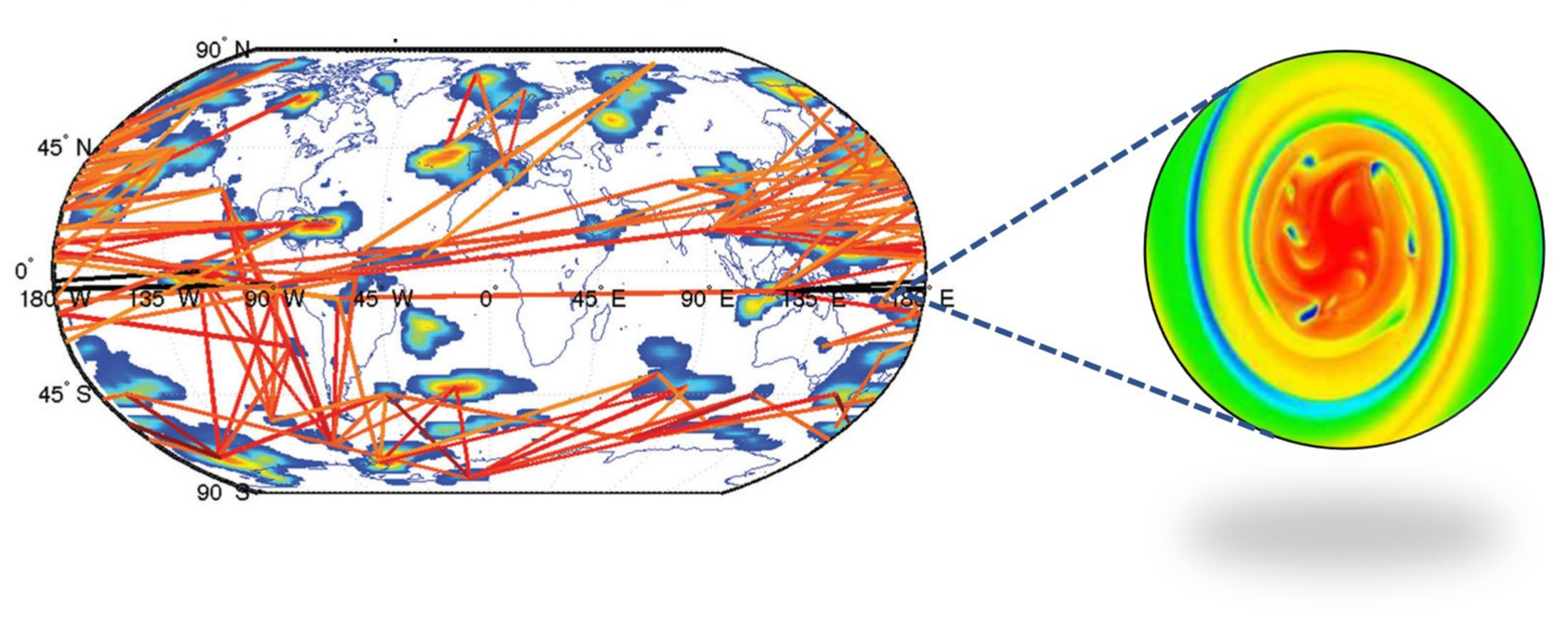}

}
\par\end{centering}
\caption{(a) Illustration of a cellular system formed by biological cells connected
by means of gap junctions to interchange chemicals and a zoom of the
internal structure of a cell. (b) Enriched conceptual system representing
the conceptual organization of animal categories. (c) Landscape ecological
system formed by patches interconnected by corridors used by some
species to move from patch to patch. Zooming-in reveals the foraging
movement of these species inside the patches. (d) Climate system formed
by a network of climatic events correlations and the internal climatic
events at local regions.}\label{fig: network picture}

\end{figure}

The first type of systems, represented in \Cref{fig: cells},
is formed by cells which are interconnected by means of their physical
contacts, such as in the case of cellular systems in tissues, neuronal
networks or astrocytic complexes, i.e., {star-shaped glial cells} 
\cite{Nedergaard}.
In this case the exo-skeleton of the system is the cellular network
\emph{per se}, and their endo-skeleton is described by the crowded environment
inside the cells, where up to 40\% of the cytoplasmic volume is occupied
by RNA, ribosomes and proteins \cite{Zhou,Zimmerman}. The situation
is similar if we consider regions inside an organ instead of individual
cells. A typical example is the consideration of anatomical or functional
regions of the brain \cite{brain_networks}. Here again the system
is characterized by an exo-skeleton formed by the brain network and
an endo-structure describing the interior of those regions. 

In the second example illustrated in \Cref{fig: circles}  we consider an enriched conceptual metaplex. This is a conceptual organization model inferred by Go$\tilde{\textnormal{n}}$i et al. \cite{semantic} from verbal fluency of 200 individuals. The study was aimed 
at finding the conceptual storage structure of the natural category of 
animals as a network. Thus, every node of the metaplex represents a 
category of animals, e.g., pets, and inside the nodes we find all the 
words in that category. 

Our third example consists of the patches formed in an ecological
landscape as illustrated in \Cref{fig: elephants}. In this case, the system of
patches and the corridors connecting them forms an exo-skeleton known
as a landscape network \cite{Urban}. The combination of the geographic
and ecological features of the individual patches determine their
endo-structures. This example will be discussed hereafter in this
work.

The final example is however at a much larger scale (see \Cref{fig: cell}).
It corresponds to a climate system in which the nodes represent geographical
regions in the world. These regions are connected by climatic correlations
or causalities, such that two regions are connected if a climatic
event in one region triggers a climatic event in the other \cite{climate_2,climate_1}. Inside every region, however, there is a vast collection
of complex phenomena taking place which are typically modeled by using
weather and local climatic models.

A common dynamical process taking place in the systems described in \Cref{fig: network picture} is diffusion, which is also ubiquitous in many other physical
\cite{Frey}, chemical \cite{Kramers}, biological and socio-economic
systems \cite{Kerner,Lax}, especially on mesoscopic scales 
\cite{Zanette}. In the systems described before, as well as
in many other complex systems, there is a trade-off between a diffusive
process taking place inside the entities of the systems and the diffusion
taking place between them. When both, the endo- and the exo-structure
of the complex system are representable by networks we can use some
of the physico-mathematical tools already available in the literature,
such as networks of networks \cite{Gao,Seshadhri} or multiplexes \cite{multiplex_1,arenasprl,multiplex_2,multiplex_notices}. However, the main challenge
here is that in many occasions the endo-structure of the system is
described by a continuous space while the exo-skeleton is a discrete
one, e.g., a network. In the case of cells, their interior is formed
mainly by cytoplasm which is formed by 80\% of water \cite{Luby-Phelps},
while their interconnection is well described by a cellular network. Go$\tilde{\textnormal{n}}$i et al. \cite{semantic} considered in their work that inside 
each category every pair of words were connected. Thus, it is assumed 
that the internal space inside categories is a continuum in which 
words are found in a random-walk-like navigation inside the memory of 
individuals.
In
the next example, where species are moving in a patchy environment,
the movement inside the patches is better described by using continuous
diffusive models  \cite{Bartumeus,Burrow}, and the hop between patches through the
narrow corridors is better accounted for by using a network-theoretic
approach \cite{Urban}. Finally, in climate systems the processes inside
the regions -- nodes -- are well described by aerodynamics and fluid
dynamics, while the causal influence between regions is well described
by a network of interactions, including long range effects \cite{climate_1,climate_2}.

The implications of this endo-exo trade-off in complex systems is
very important for understanding their functioning. There is experimental
evidence, for instance, that the movement of small molecules inside
a cell follows an anomalous diffusive processes, either subdiffusive
\cite{Golding,Nicolau,Weber} or superdiffusive \cite{Reverey}.
The distinction between normal (or Fickian) and anomalous diffusion is made on the basis of the scaling of the mean squared displacement $\langle x^2\rangle=\langle (x-x_0)^2\rangle$
of the diffusive particle with time \cite{Bouchaud,Metzler_1}, where $x$ is the current position of a particle and $x_0$ is the initial position. See Supplementary Information for a more detailed description of anomalous diffusion.
While for normal diffusion the mean squared displacement scales linearly with time $\langle x^2\rangle \sim t$, for anomalous diffusion it scales as a power-law $\langle x^2\rangle \sim t^\alpha$, with exponent $\alpha\in(0,2)$ larger
(super-) or smaller (sub-) than one. This
means that on long time scales in a superdiffusive process the space explored by
the diffusive particle is larger than the one explored in
a normal diffusion in exactly the same time, due to a high probability of long range hopping.  In the case of a small molecule inside a cell, the anomalous
diffusion is mainly due to the crowded
environment inside the cells \cite{Sanabria,Smith}. However, there
are more complex mechanisms inside certain types of cells that can lead to anomalous
diffusive processes. An example is the generation of
calcium waves \cite{Chen} which have been observed in cardiac muscle
\cite{Fabiato}, in skeletal muscle fibre \cite{Endo}, medaka eggs \cite{Ridgway}
and astrocytes \cite{Cornell-Bell}. This subdiffusive $\textnormal{Ca}^{2+}$
movement may be involved with cardiac  \cite{Lakatta,Lu} and brain diseases
\cite{Wetherington,Jo}.

When global systems, such as tissues \cite{diffusion_brain} or
the whole brain \cite{brain_networks}, are analyzed, superdiffusive
processes have been experimentally observed for fresh specimen of
carcinoma, fibrous mastopathies, adipose and liver tissues \cite{Kopf}
as well as for signals navigating across regions of the brain \cite{costa2016foraging}.
Interesting research questions emerge from these experiments: Is the anomalous diffusive
behavior of a global system the consequence of sub or superdiffusive
processes inside the entities, e.g., cells or regions? Are they the
result of the anomalous diffusion between the entities only? Is it
the combination of the exo- and endo-structures which determines the nature of the diffusive process in a complex system? How relevant is the endo-structure depending on the type of network, i.e. small-world network? Similar questions emerge for the
analysis of other systems, such as the 
landscape system discussed
above. 
In landscape systems the superdiffusive movement of species inside
a patch is well documented and described by continuous models \cite{Bartumeus,Burrow,Viswanathan}.
Can this behavior alone determine the nature of the diffusive process
at the global landscape level? \\
\begin{wrapfigure}{r}{5cm}
\begin{centering}
\includegraphics[scale=0.5]{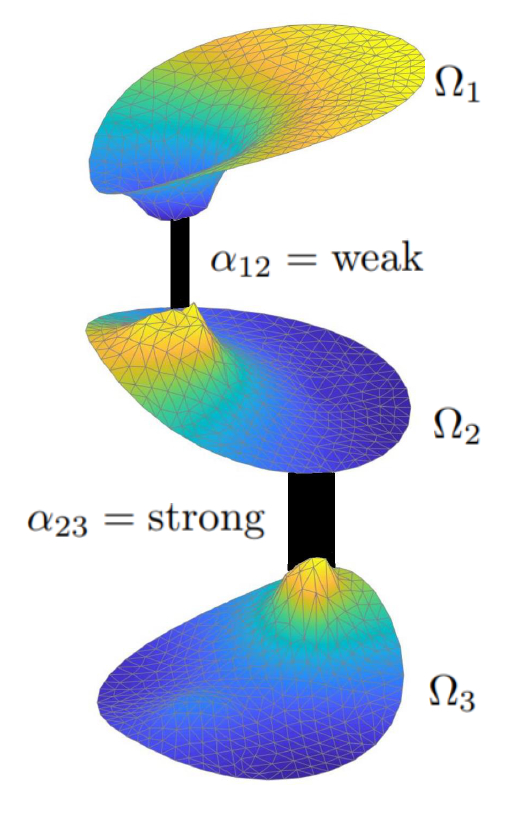}
\par\end{centering}
\caption{Diffusion in a metaplex. The domains $\Omega_j$ are the nodes of the graph (exo-skeleton) in which a continuous diffusion process takes place. The connections with different strengths, given by $\alpha_{12}$ and $\alpha_{23}$, account for the edges of the graph. The colours indicate the "density" of the diffusing particles, with yellow for higher concentration.}
\label{FigureMetaplex}
\end{wrapfigure}

In this work we answer these questions by introducing the concept
of \textit{metaplexes}. Informally, a metaplex is a representation
of a complex system in which the internal structure of the nodes
and the interconnection between them are considered at the same time
(see \Cref{FigureMetaplex}). Thus, a dynamical system on a metaplex consists of the
coupling between the dynamics inside the entities, typically a continuous
space, and the dynamics between these entities which is controlled
by the inter-entity connectivity. 

Our results show that in a linear metaplex, superdiffusion due to long range hopping in a network, as in \cite{estradalaa}, survives  irrespective of the internal structure of the nodes. On the other hand, we prove that superdiffusion in the nodes can speed up regular diffusion in the metaplex, but it cannot lead to superdiffusion.  The geometry of the nodes and their coupling play crucial roles for the global dynamics, which we explain in a combination of analysis and numerical experiments.
The results shed light on the rich  and substantially different nature of the dynamics of metaplexes and the interplay of their exo- and endo-structure, also in the real-world systems considered in \Cref{superdiffsection}.

Here we start by a formal definition of a metaplex
and a dynamical system on it. We then study a linear toy model that allows
us to understand some of the fundamental principles of diffusive processes
on metaplexes. Finally, we study two real-world metaplexes
pointing at the potential applications of this representation
of complex systems.

\section{Preliminaries}\label{sec: def}

In this section we define the concepts and settle the notation to
be used in this work. Here we consider two kinds of diffusive dynamical
systems, one taking place in a continuous space and the other in a
discrete one. Let us start by defining the diffusive process on the
continuous space. 

Let $u\left(t,\mathbf{x}\right)$ be the density
of the diffusive particle at time $t$ for $\mathbf{x}$ in a domain
in $\mathbb{R}^{n}$. The density $u\left(t,\mathbf{x}\right)$ evolves
according to

\begin{equation}
\partial_{t}u\left(t,\mathbf{x}\right)=\left(-\Delta\right)^{s}u\left(t,\mathbf{x}\right),
\end{equation}
where $\left(-\Delta\right)^{s}$ denotes a \textit{fractional
Laplacian operator} for $s\in\left(0,1\right].$  The application of
the fractional Laplacian to $u\left(t,\mathbf{x}\right)$ on $\mathbb{R}^{n}$
at a fixed time $t$ is given by

\begin{equation}\label{e:fl}
(-\Delta)^{s} u (\mathbf{x}) = c_{n,s}\, P.V. \int_{\mathbb{R}^n} \frac{u(\mathbf{x})-u(\mathbf{y})}
{\vert \mathbf{x} - \mathbf{y} \vert^{n+2s}} d\mathbf{y} = c_{n,s} \lim_{\varepsilon \to 0^+} 
\int_{\{|\mathbf{y}|>\varepsilon\}} \frac{u(\mathbf{x})-u(\mathbf{y})}{
\vert \mathbf{x} - \mathbf{y} \vert^{n+2s}} d\mathbf{y}\ ,
\end{equation}
where $P.V.$ denotes the Cauchy principal value, and $c_{n,s}$ is
the following normalization constant in terms of Euler Gamma functions
\[
c_{n,s} = \dfrac{2^{2s} s \Gamma\left(\frac{n+2s}{2}\right)}{\pi^{\frac{n}{2}}
\Gamma\left(1-s\right)}.
\]

In a domain $\Omega\subset\mathbb{R}^{n}$ it is better to define the
fractional Laplacian in terms of the following bilinear form:

\begin{equation}
\int_{\varOmega}\left(-\Delta\right)^{s}u\left(\mathbf{x}\right)v\left(\mathbf{x}\right)d\mathbf{x}=\dfrac{c_{n,s}}{2}\iint_{\left(\Omega\times\mathbb{R}^{n}\right)\cup\left(\mathbb{R}^{n}\times\Omega\right)}\dfrac{\left(u\left(\mathbf{x}\right)-u\left(\mathbf{y}\right)\right)\left(v\left(\mathbf{x}\right)-v\left(\mathbf{y}\right)\right)}{\left|\mathbf{x}-\mathbf{y}\right|^{n+2s}}d\mathbf{y}d\mathbf{x},
\end{equation}
for $u,v$ belonging to the Sobolev space $H^{s}\left(\Omega\right)$.
When $s=1$ this operator corresponds to the differential Laplacian
operator $\Delta u=\partial_{x_{1}}^{2}u+\cdots+\partial_{x_{n}}^{2}u$
\cite{Metzler_1, Metzler_2} with Neumann boundary conditions \cite{dipierro2017nonlocal}.

Let us now move on to the diffusive process on a discrete space. Let
$G=\left(V,E\right)$ be a simple, connected graph with $\left|V\right|=N$
nodes and $\left|E\right|=m$ edges. The degree of a  node in $ V$ is the number of its nearest neighbors.  Let $\textnormal{dist}\left(v,w\right)$ be
the shortest path distance between the nodes $v\in V$ and $w\in V$,
i.e., the number of edges in one shortest path connecting both nodes.
Let $d_{max}=\max_{v,w}\textnormal{dist}\left(v,w\right)$ be the maximum distance
between any pair of nodes in $G$, i.e., the graph diameter. 

Hopping between the nodes in the discrete
space of a network occurs through sinks (dark blue area in \Cref{FigureMetaplex}) and sources (yellow area), where particles are allowed to hop to non-nearest nodes with a certain probability. Since the metaplexes that we study are undirected, the sinks also act as sources and vice-versa (see Supplementary Note 2). The
evolution of the density of the diffusive particle among the nodes
of the graph is controlled by the generalized diffusion equation \cite{estrada2012path}:

\begin{equation}
\dfrac{d\mathbf{u}\left(t\right)}{dt}=-\left(\sum_{d=1}^{d_{max}}c_{d}\Delta_{d}\right)\mathbf{u}\left(t\right),\qquad\mathbf{u}\left(0\right)=\mathbf{u}_{0},
\end{equation}
where $\Delta_{d}$ is the $d$-path Laplacian operator of the graph
transformed by the coefficients $c_{d}$. This operator is defined
as follows. Let $C\left(V\right)$ denote the set of complex-valued
functions in $V.$ Let $f\in C\left(V\right),$ then \cite{estrada2012path}

\begin{equation}
\Delta_{d}f\left(v\right)=\sum_{w\in V,\textnormal{dist}\left(v,w\right)=d}\left(f\left(v\right)-f\left(w\right)\right),\qquad v\in V.
\end{equation}

Notice that if we consider only $\textnormal{dist}\left(v,w\right)=1$ in the above
definition, this operator becomes the graph Laplacian $\Delta_{1}=L=K-A$, where $K$ is a diagonal matrix of node degrees and $A$ is
the adjacency matrix of the graph.
The transformation of the $\Delta_{d}$ operators through the coefficients
$c_{d}$ is aimed at tuning the hopping of the diffusive particle
between nearest and non-nearest neighbors in the graph. For instance,
let

\begin{equation}\label{kpathmel}
\Delta_{M}f:=\sum_{d=1}^{d_{max}}d^{-s}\Delta_{d}f,
\end{equation}
designate the Mellin-transformed $d$-path Laplacians of the graph where $d_{max}$ is the diameter of the graph..
Then, when $s\rightarrow\infty$ the transformed operator tends to
$\Delta_{1}=L=K-A.$ However, when $s$ is relatively small the diffusive
particle can hop not only to nearest neighbors but also to distant
nodes with certain probability that decays with the shortest path
distance separating the origin and destination of the particle. It
was recently proved that in the one-dimensional lattice when $s\in\left(1,3\right)$
there is a superdiffusive process on the graph \cite{estradalaa}.
The same happens for two-dimensional lattices when $s\in\left(2,4\right)$
\cite{estrada2018path}. Other transforms,
such as the 
exponential one, also known as the Laplace transform:

\begin{equation}\label{kpathlap}
\Delta_{e}f:=\sum_{d=1}^{d_{max}}e^{-ds}\Delta_{d}f,
\end{equation}
can also be used in the generalized diffusion equation but they have
been proved to show no superdiffusive processes on the graph \cite{estradalaa}.

In the analysis of the diffusion processes on metaplexes we will refer
to the rate at which the diffusive particle reaches the steady state $\bar{u}$ as $t\rightarrow \infty$. The rate of convergence of the diffusion on the metaplex is controlled by the standard deviation of the density, $\sigma\left(t\right)=\sqrt{\dfrac{1}{N-1}\sum_{i}\left(\int_{\Omega_i}u_{i}\left(t,\mathbf{x}_i\right)-\bar{u}_i\right)^{2}}$, 
where the summation is over all $N$ nodes in $ V$. In our examples the 
steady state $\bar{u}$ is the uniform distribution on $V$. 

 
\section{Metaplexes: Structure and Dynamics}\label{sec: diffusion}
We have previously defined informally what a metaplex is. Here we
 present a formal definition of a metaplex and of a dynamical system
on it. Let us start with the following.

\begin{definition}
A metaplex is a $4$-tuple $\Upsilon=\left(V,E,\mathcal{I},\omega\right)$,
where $(V,E)$ is a graph, $\omega=\{\Omega_{j}\}_{j=1}^{k}$ is a
set of locally compact metric spaces $\Omega_{j}$ with Borel measures
$\mu_{j}$, and $\mathcal{I}:V\to\omega$.
\end{definition}

For instance, let us consider a simple metaplex in which $(V,E)$
is a path graph, e.g., the path graph of $3$ nodes with each node
given by the unit disk $B(0,1)\subset\mathbb{R}^{2}$. In this case
$\omega=\{B(0,1)\}$ and $\mathcal{I}$ is constant. The resulting
metaplex is illustrated in \Cref{FigureMetaplex}. The endo-structure of this
metaplex is given by the internal structure inside the unit disk,
and its exo-structure is given by the connectivity of the three nodes
in the form of a path. In the current work we focus only on metaplexes
in which the internal structure of the nodes is continuous. That is,
when $\omega$ is a set of open domains $\Omega_{j}\subset\mathbb{R}^{n}$,
each endowed with the Lebesgue measure. Other scenarios in which the
internal structure is a discrete space, not only a graph, should
be considered in a separate analysis, and they will
not be treated in the current work. Now, we are in condition to define
a dynamical system on a metaplex.

\begin{definition}\label{defdyn}
A dynamical system on a metaplex $\Upsilon=\left(V,E,\mathcal{I},\omega=\{\Omega_{k}\}\right)$
is a tuple $(\mathcal{H},\mathcal{T})$. Here $\mathcal{H}={\left\{ H_{v} : L^{2}(\Omega_{\mathcal{I}(v)}, \mu_{\mathcal{I}(v)}) \to L^{2}(\Omega_{\mathcal{I}(v)}, \mu_{\mathcal{I}(v)})\right\} _{v\in V}}$
is a family of operators such that the initial value problem $\partial_{t}u_{v}=H_{v}(u_{v})$,
$u_{v}|_{t=0}=u_{0}$, is well-posed, and $\mathcal{T}=\{T_{vw}\}_{(v,w)\in E}$
is a family of bounded operators $T_{vw}:L^{2}(\Omega_{\mathcal{I}(v)},\mu_{\mathcal{I}(v)})\to L^{2}(\Omega_{\mathcal{I}(w)},\mu_{\mathcal{I}(w)})$.
\end{definition}

There are many possible dynamical systems that can be considered on
a metaplex under the previous general definition. In the current work
we focus only on the study of diffusive processes at the endo- and
exo-structure of the metaplexes. In this scenario we consider a continuous
diffusion equation inside the nodes where the evolution of the density
of the diffusive particle is controlled by the fractional Laplacian
operator as described in \Cref{sec: def}. At the exo-skeleton we consider
a diffusive process in which the particle hops from a node to another
as controlled by the $d$-path Laplacian of the graph also described in
\Cref{sec: def}. The in and out motion of the diffusive particle between
nodes is carried out through the so-called\textit{ sources} and \textit{sinks}
inside the nodes. A source is a subdomain inside the node from which
a diffusive particle can emerge to the interior of the node. A sink
is another (not necessarily different) subdomain inside the node from
which a diffusive particle can abandon the interior of the node towards
another node in the metaplex.

In the simplest case, diffusive particles move between different nodes
through such sinks and sources in the interior, corresponding to a
coupled system of diffusion equations for the density $u_{j}(t,\mathbf{x})$
of particles in the node $v_{j}\in V$:
\begin{align}\label{sourcecoupling}
\partial_t u_j(t,\mathbf{x}) &= \mathrm{div}J_j(u_j(t,\mathbf{x})) -\sum_{i : (v_j, v_i) \in E} \alpha_{ij}(\mathbf{x})u_j(t,\mathbf{x})\\ & \qquad +  \sum_{i : (v_i, v_j) \in E} \alpha_{ji}(\psi_{ji}^{-1}(\mathbf{x})) \mathrm{det}(\nabla \psi_{ji}^{-1})u_i(t,\psi_{ji}^{-1}(\mathbf{x})),\nonumber
\end{align}
for $(t,\mathbf{x})\in(0,\infty)\times\Omega_{j}$. Here $J_{j}$
is the flux of particles, and $\mathrm{div}J_{j}$ is the generator
of the diffusion process in $\Omega_{j}$, for example $\mathrm{div}J_{j}=\Delta$
in the case of a normal diffusion, while $\mathrm{div}J_{j}=-(-\Delta)^{s}$
for the superdiffusive L\'{e}vy process. The edges $(v_{i},v_{j})\in E$
are realized by a map $\psi_{ji}:\Omega_{j}\to\Omega_{i}$ which specifies
the jumps between domains, and the coefficients $\alpha_{ji}(\mathbf{x})$
are transition rates from $\Omega_{i}$ to $\Omega_{j}$: Particles
jump from $\mathbf{x}$ to $\psi_{ji}(\mathbf{x})$ with amplitude
$\alpha_{ij}(\mathbf{x})$ and from $\psi_{ji}(\mathbf{x})$ to $\mathbf{x}$
with amplitude $\alpha_{ji}(\psi_{ji}^{-1}(\mathbf{x}))$. Physically,
the system \cref{sourcecoupling} arises for nodes which correspond
to spatially distant domains.

{For the system of diffusion equations \cref{sourcecoupling}, $H_{v_j} u_j = \mathrm{div} J_j(u_j)$, while $T_{vw}$ correspond to the entries of the  transition matrix of the network given by the functions $\alpha_{ij}$.}


{Diffusion in a network may be studied from its generator, the graph Laplacian $\Delta_1$. A dynamical system in a metaplex is similarly described by an operator matrix $\mathcal{D}$. For the diffusion processes like \eqref{sourcecoupling} it takes the abstract form }
{\begin{equation}\label{eq: diffeq}\partial_t u = \mathcal{D}u\ .\end{equation}}
{$\mathcal{D}$ is an $N\times N$} block operator matrix of unbounded operators on the product space {$\bigoplus_{j=1}^N L^2(\Omega_j, \mu_j)=L^2(\Omega_1,\mu_1)\times \dots \times L^2(\Omega_N,\mu_N)$}, {with $N$ the number of nodes in $V$}. 
 In line with \Cref{defdyn} we write $\mathcal{D} = H+T$, where $$H=
\begin{pmatrix}
\mathrm{div} J_1 & 0& 0& \cdots & 0\\
0 & \mathrm{div} J_2 &0& \cdots & 0\\
\vdots & 0 & \ddots& & \vdots\\
\vdots & \vdots & & \ddots& \vdots\\
0&0& \cdots & \cdots & \mathrm{div} J_N \\
\end{pmatrix}  \ ,
$$
{is given by the operators $H_{v_j} u_j = \mathrm{div} J_j(u_j)$ and}
$$
T=
\begin{pmatrix}
   \alpha_{11} T_{11}& \alpha_{12}T_{12}& \cdots & \alpha_{1N}T_{1N}\\
\alpha_{21}T_{21} &    \alpha_{22}T_{22}& \cdots & \alpha_{2N}T_{2N}\\
\vdots & \vdots &\ddots & \vdots\\
\alpha_{N 1}T_{N 1}&\alpha_{N 2}T_{N 2} & \cdots &   \alpha_{NN}T_{NN}\\
\end{pmatrix}\ ,
$$
{describes the network diffusion defined a $\mathcal{T}$.} Provided the adjoint $T^*\begin{pmatrix}
           1 \\
           1 \\
           \vdots \\
          1
         \end{pmatrix}=0$, particles are conserved.
Here the $T_{ij}$ are transition operators between $\Omega_i$ and $\Omega_j$, as given by the sources and sinks, and $\alpha_{ij}$ {are} the transition probabilities. 

{For a network, spectral properties of the network Laplacian determine the long-time behavior of diffusion. Analogously, the long-time behavior of the linear diffusion equation \eqref{eq: diffeq} in the metaplex is determined by the spectral properties of $\mathcal{D}$ \cite{tretter}.} While the spectrum of $H$ is determined by the internal structure of the nodes, the spectrum of $T$ combines the details of the location and strength of sinks, respectively sources, with the external network structure.\\


{In addition to the coupling by sinks and sources as in \eqref{sourcecoupling}, numerous other types of couplings can be considered within the framework of metaplexes. While their detailed analysis is beyond the scope of this paper, we mention two important examples.} 

{The first example is a special case of \eqref{sourcecoupling}, with all nodes of the same geometry $\Omega = \Omega_j$ for every $j$ and local transitions $\psi_{ji} = \mathrm{Id}$, where $\textnormal{Id}$ is the identity operator (matrix)
of the appropriate dimension. In this case the} density $u$ may be interpreted as a vector valued function $u = (u_1, \dots,u_N): [0,\infty)\times \Omega \to \mathbb{R}^N$, and the network encodes the dynamics  of the ``internal state'' of the particle described by a vector in $\mathbb{R}^N$. In biology, such processes are of interest to describe the diffusion of complex organisms \cite{perthame2018fractional,erban2005signal,erban2004individual}. 

In complex systems such as road systems or the transport of chemicals between cells, the coupling between the nodes occurs through the boundary {$\partial\Omega_j$}, {not through internal sinks and sources}. Every edge $(v_i, v_j) \in E$ is physically realized by open entrances and exits $\Gamma_{ij} \subset \partial \Omega_i$ and $\Gamma_{ji} \subset \partial \Omega_j$, together with a homeomorphism $\phi_{ij} : \Gamma_{ij} \to \Gamma_{ji}$ identifying points between them. We define $\Gamma_{i0} = \partial \Omega\setminus \bigcup_{j} \Gamma_{ij} $. If there is an edge between $\Omega_i$ and $\Omega_j$, particles leave $\Omega_i$ through $\Gamma_{ij}$ and arrive at $\Gamma_{ji}$ in $\Omega_j$.

The resulting system of diffusion equations is coupled through the boundary conditions, with a Kirchhoff's law: For $\mathbf{x} \in \bigcup_i \Gamma_{ji}$
$$u_j(t,\mathbf{x}) = \sum_{i: \mathbf{x} \in \Gamma_{ji}} \alpha_{ji} \mathrm{det}(\nabla_{\partial \Omega}\phi_{ji}) u_i(t,\phi_{ji}(\mathbf{x})) \ ,$$
$$J(u_j(t,\mathbf{x})) \cdot \nu_j(\mathbf{x}) = -\sum_{i: x \in \Gamma_{ji}} \alpha_{ji} \mathrm{det}(\nabla_{\partial \Omega}\phi_{ji}) J(u_i(t,\phi_{ji}(\mathbf{x}))) \cdot \nu_i(\phi_{ji}(\mathbf{x})) \ ,$$
where $\nu_j$ is the exterior unit normal vector.
The number of particles is preserved if the transition probabilities satisfy $\sum_{i}\alpha_{ji} = 1$ for every $j$.

\section{Operators and spectra for metaplexes}\label{sec: operators}


\subsection{Variation of eigenvalues with fractional exponent and spatial scale}\label{sec: variation of eig}

Note that $H = \bigoplus_{j=1}^N H_j=\bigoplus_{j=1}^N \mathrm{div} J_j$ is a block diagonal operator matrix acting on $L^2(\Omega_1) \times \cdots \times L^2(\Omega_N)$. A basis of eigenfunctions of $H$ is constructed from bases $\{u_{j,k}\}_{k}$ of eigenfunctions of $\mathrm{div} J_j$ in $\Omega_j$ and it is given by $\{u_{j,k} e_j \}_{j,k}$ where the subscripts  denotes the $k$-th eigenfunction in the domain $\Omega_j$. Here $e_j$ denotes the $j$-th standard unit vector in $\mathbb{R}^N$.

If $H_j=-(-\Delta)^{s_{nod}}$ is the fractional Laplacian in $\Omega_j$ with Neumann boundary conditions, the eigenvalues of $H_j$ are homogeneous functions of the spatial scale, i.e., if the spatial variable is scaled, then also the function, $f(\alpha x)=\alpha^px$, where $p$ is the degree of homogeneity. More precisely, the bilinear form of $H_j$ is homogeneous under scaling $\Omega_j \mapsto  \Lambda \Omega_j$, $\Lambda >0$. For instance, if $\Omega_j=B(0,5)$ is a ball of radius $5$, then the scaling of the domain can be interpreted as $\Lambda\Omega_j=B(0,\Lambda 5)$. 

From their characterization in terms of a Rayleigh quotient, the eigenvalues in $\Lambda \Omega_j$ are given by $\{\Lambda^{-2s_{nod}} \lambda_{j,k}\}_{k=1}^\infty$, if $\{\lambda_{j,k}\}_{k=1}^\infty$ are the eigenvalues of $H_j$ in $\Omega$ \cite{courant2008methods}. Only the lowest eigenvalue $\lambda_{j,1} = 0$ is fixed under this scaling.

 We see that for small domains, $\Lambda \to 0$, the spectral gap $\lambda_{j,2}-\lambda_{j,1}$ in the node $\Omega_j$ increases with $s_{nod}$, and therefore Brownian motion in the nodes gives the fastest convergence to equilibrium. For large domains, $\Lambda \to \infty$, the spectral gap $\lambda_{j,2}-\lambda_{j,1}$ in the nodes $\Omega_j$ decreases with $s_{nod}$, and therefore the long jumps of the fractional diffusion lead to faster equilibration than Brownian motion inside each node.
 
In a further section (see \Cref{sec: numerics}) we will illustrate computationally these findings.



\subsection{Weakly and strongly interacting limits}\label{weakstrongsect}

The spectral properties of $\mathcal{D}_\varepsilon=H+\varepsilon T$ are most easily understood in the limit $\varepsilon \rightarrow 0$ of weak network interactions between the nodes, respectively the strong network interactions for $\varepsilon \rightarrow \infty$.   In the former case,  a particle is trapped for a long time inside the domain in which it was initially placed. The time scale of the slow global equilibration is determined by $T$. In the latter case, the transient dynamics is governed by the dynamics of the network and geometry of the sinks and sources, i.e. $T$, but the dynamics $H$ inside the nodes determines the long-time approach to equilibrium.



From standard perturbation theory, for general, symmetric interactions $T$ the stationary states corresponding to $\lambda_{j,1} = 0$, $j\in\{1,\dots,N\}$, split into eigenvalues
$\varepsilon \widetilde{\lambda}_k +o(\varepsilon)$ according to the eigenvalues $\widetilde{\lambda}$ of $T$ restricted to the kernel of $H$. The kernel of $H$ is spanned by the constant functions $1$ in $\Omega_j$, $j=1, \dots, N$. Hence, $(T|_{\mathrm{ker}\ H})_{ij} =\frac{1}{\sqrt{|\Omega_i| |\Omega_j|}}\int_{\Omega_j} \alpha_{ij} T_{ij}1\ dx$, $i,j = 1,\dots, N$, is given by an effective graph Laplacian for the metaplex, in which the internal structure has been integrated out. The eigenvalues $\widetilde{\lambda}_k$ of this matrix determine the spectral gap $\varepsilon(\widetilde{\lambda}_2-\widetilde{\lambda}_1)$ in terms of the spectral gap of the effective graph Laplacian $T|_{\mathrm{ker}\ H}$, independent of $H$. Higher eigenvalues will depend on the location of the sources and sinks and the eigenfunctions of $H_j$. 

Multiple eigenvalues arise, in particular, when several of the $H_j$ coincide. In this case the eigenvalues are determined by the restriction $T_r$ of $T$ to the subnetwork of those nodes where the internal diffusion has the eigenvalue $\lambda_{j,k}$. 

Higher eigenvalues depend on the location of the sources and sinks. The spectral gap of the network is $\varepsilon \lambda_2$, and thus independent of the diffusion process $H$. 

Local equilibration within the nodes, however, happens on faster time scales, with little effect of the coupling. For small $\varepsilon$, the gap between the higher bands $\lambda_{j,k} + \varepsilon \lambda_n^r$ to the equilibrium is determined by the diffusion $H$, up to terms of order $\varepsilon$, where  $\{\lambda_k^r\}$ are the eigenvalues of $T_r$.


 
 The case where  $T_{ij} =  \mathrm{Id}$ is discussed in {the Supplementary Material} (Note 3).
For general $T_{ij}$ the spectrum of the network interaction operator $T$ will {be considered elsewhere.}

\subsection{Numerical results of the spectral properties}

We illustrate the spectral gap of $\mathcal{D}_{\varepsilon}$ for a linear network of $11$ unit disks $\Omega =\Omega_j = B(0,1)\subset \mathbb{R}^2$. The lowest $10$ nonzero eigenvalues are considered as a function of $\varepsilon$, the coupling operator $T$ and the diffusion $H$. We choose the generator $H_j$ of the diffusion process inside $\Omega_j$ as a fractional Laplacian $-(-\Delta)^{s_{nod}}$ with L\'{e}vy exponent $s_{nod}$ and approximate it by finite elements on quasi-uniform spatial meshes. See \Cref{fig: mesh}  below for a plot of the mesh and \cite{m3as} for the approximation of the fractional Laplacian. 

The network coupling $T_{ij}$ is taken to be a $d$-path Laplacian \eqref{kpathlap} with hopping between $\Omega_i$ and $\Omega_j$ proportional to an exponential $2^{-s_{net}|i-j|}$. 

As $H$ is unbounded the eigenvalues of its discretization extend over several orders of magnitude, while the spectral gap is tiny. To resolve it, the discretization of the mesh $h$ needs to be small compared to the strength of the coupling, and standard Matlab routines do not identify the bounded branches of eigenvalues obtained in \Cref{weakstrongsect} for large $\varepsilon$. Discretization errors significantly increase as $s_{nod} \to 0$ because the smoothness of the solutions in $\Omega$ decreases. 

\begin{figure}
    \centering
    \includegraphics[scale=0.4]{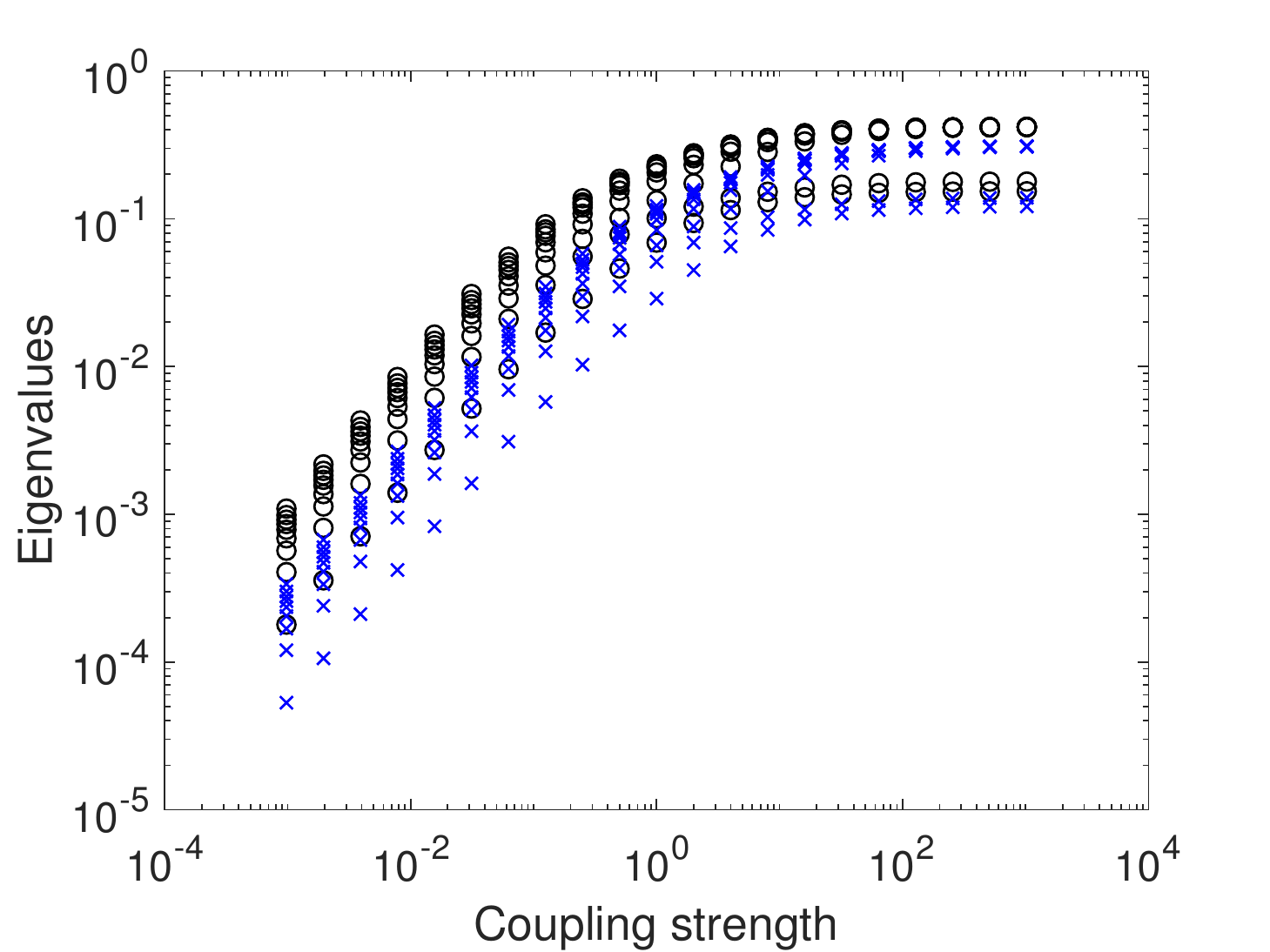}
    \caption{ Comparison of eigenvalues for the case of different discretizations $h$ of the mesh: a coarse mesh (347 degrees of freedom) $(\circ)$ and a finer mesh (1325 degrees of freedom) $(\times)$. Here $s_{net}=0.8$ and $s_{nod}=0.4$ and we consider different coupling points.}
    \label{fig: comparison between meshes}
\end{figure}

\cref{fig: comparison between meshes} indicates the complications inherent in approximating the spectrum of a large system of differential operators. The lowest $10$ nonzero eigenvalues of $\mathcal{D}_\varepsilon$ are depicted for the exponential coupling as a function of the coupling strength $\varepsilon \in [2^{-10},2^{10}]$ for meshes of $347$ and $1325$ degrees of freedom, $s_{net}=0.8$, $s_{nod}=0.4$. We consider a prototypical metaplex coupling as depicted in \cref{FigureMetaplex} for $\alpha_{ij}=10$. While results agree for large coupling strengths, the spectral gap is significantly smaller for the finer mesh at small coupling. Nevertheless, the qualitative behavior from \Cref{weakstrongsect} is recovered: the gap increases linearly for small $\varepsilon$, and for large $\varepsilon$ it converges to the lowest eigenvalue of $H$. 

\begin{figure}[!ht]
\includegraphics[scale=0.4]{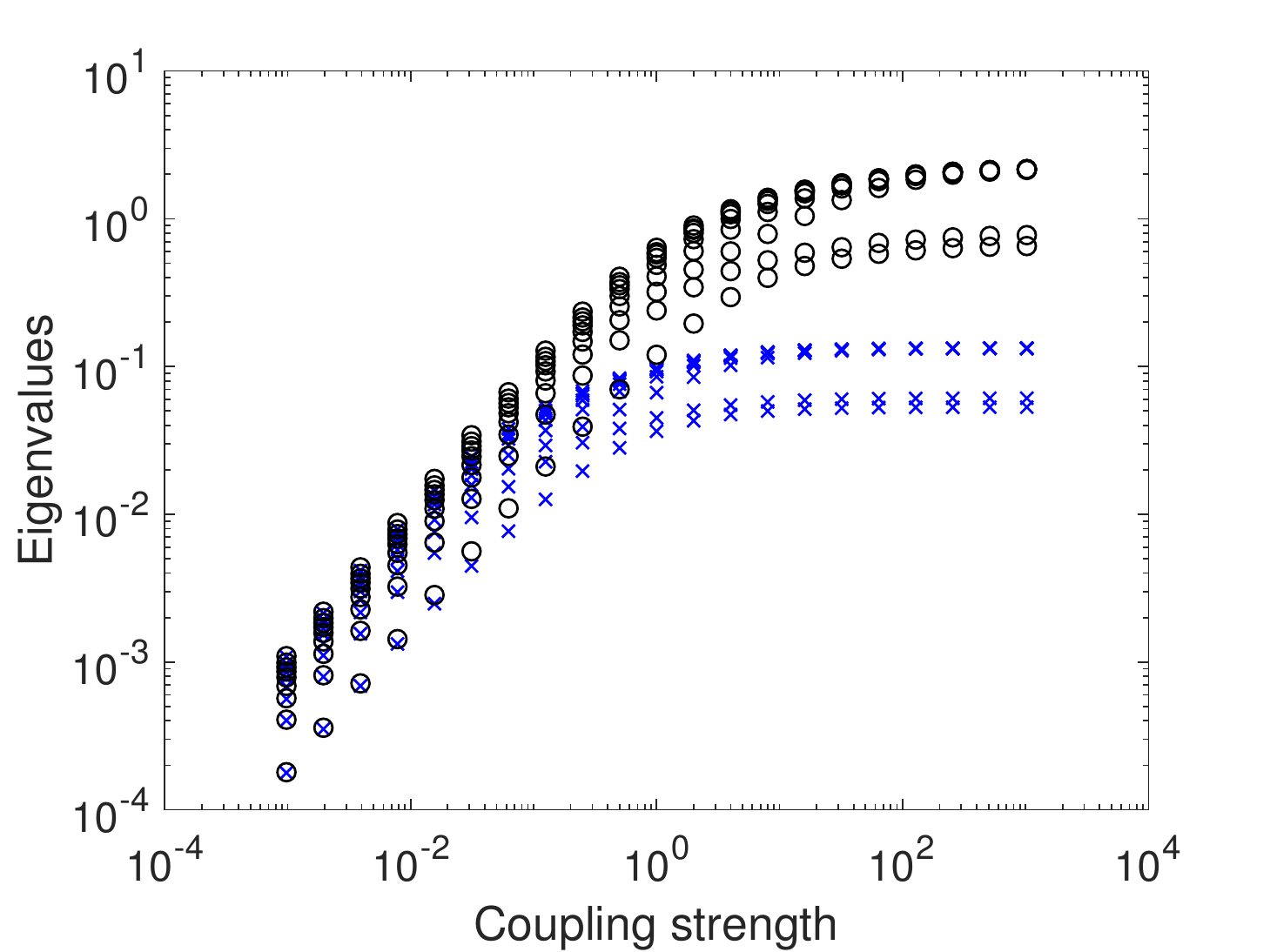}\includegraphics[scale=0.4]{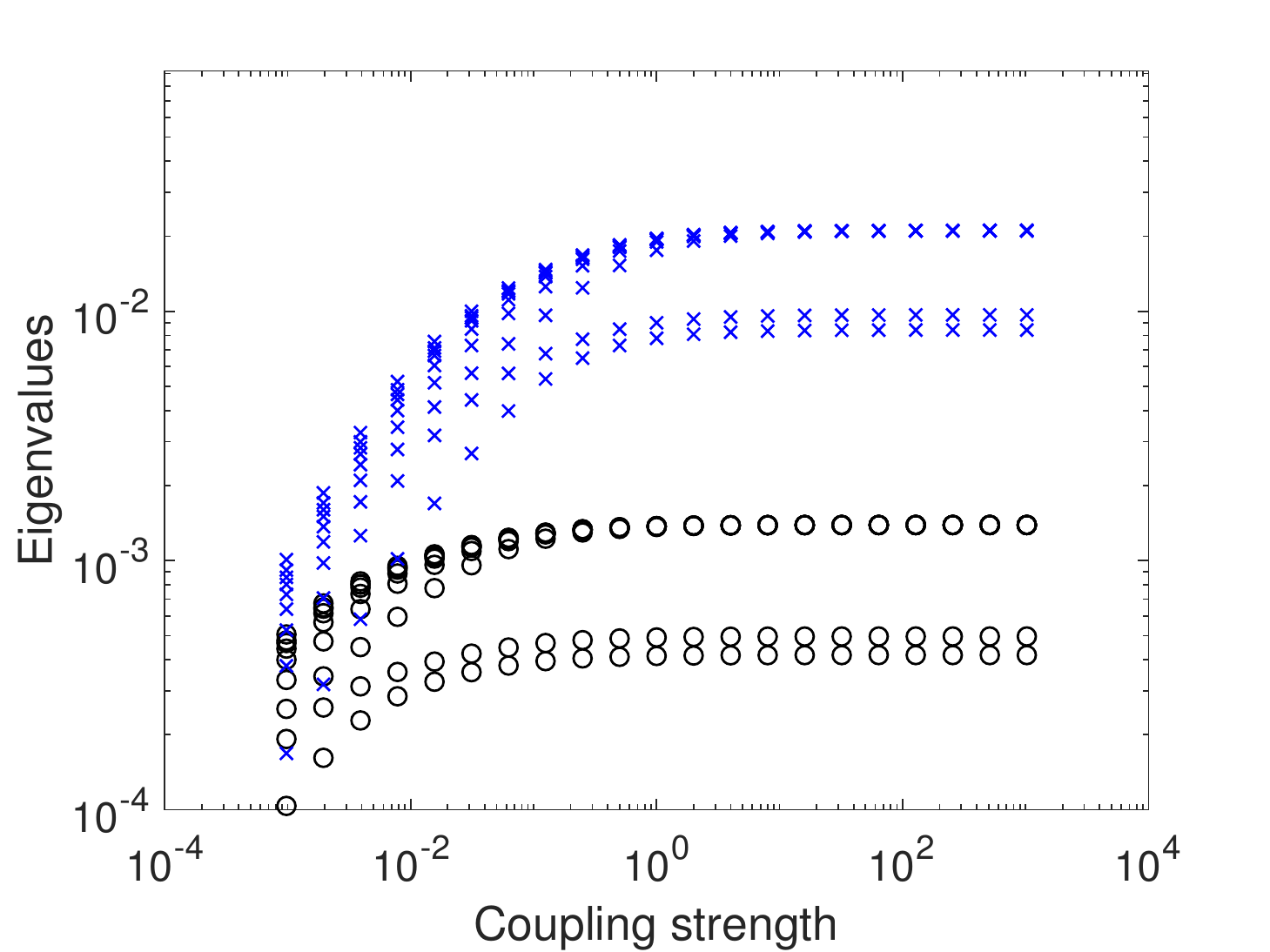}
\caption{ Eigenvalues for small (left) and big (right) nodes. In both cases $s_{net}=0.8$ with $s_{nod}=0.2$ (crosses) and $s_{nod}=0.8$ (circles).}\label{doweneedthis}
\end{figure}

\cref{doweneedthis} connects the theory of \Cref{weakstrongsect} with the numerical results from \Cref{sec: numerics} and evidences the effect of the geometry of the nodes on the dynamics of the network. In the case of the small nodes (\cref{doweneedthis} left) when the coupling is weak, the dynamics of the network is dominated by the diffusion process inside the nodes and we have almost no hopping between domains. As we increase the coupling strength, for $s_{nod}=0.8$ the spectral gap is bigger than that for $s_{nod}=0.2$ in agreement with \Cref{sec: variation of eig}. Similar to the case of multiplex networks, the first non-zero eigenvalue of $\mathcal{D}_\varepsilon$ is related to the equilibration time in the whole networks, i.e., $t=\lambda_2^{-1}$ \cite{arenaspre}. From \cref{doweneedthis} (left) the equilibration time for the case $s_{nod}=0.2$ is much longer than for the normal diffusion case inside the nodes. 
In a further section of this work (see \Cref{sec: numerics}) we obtain 
computational results that confirm the current findings.


In the case of big nodes (\cref{doweneedthis} right) the spectral gap decreases as we increase the parameter $s_{nod}$. This means that, opposite to the small node case, a metaplex with superdiffusion inside the nodes ($s_{nod}=0.2$) reaches the equilibrium faster than for internal Brownian motion. 

\section{Analysis of diffusion in metaplexes}\label{sec: analysis of diffusion in metaplexes}

In this section we indicate how PDE techniques allow an analysis of the time-dependent diffusion introduced in \Cref{sec: diffusion} in a metaplex of bounded domains $\omega = \{ \Omega_j \subset \mathbb{R}^n\}_{j=1}^k$,  as illustrated in \Cref{FigureMetaplex}. We compare the interaction of diffusion, respectively superdiffusion, in the nodes with either short or long range coupling in the external network. While for general complex networks the connectivity and large scale geometry of the network will crucially influence the dynamics, here we focus on simple linear metaplexes $V = \mathcal{Q}= \{\dots, -2, -1, 0,1,2,\dots\}$ and short times. 

More precisely, we assume that the diffusion process in each node $\Omega_j$ is governed by either a fractional Laplacian $H_j = -(-\Delta)^{s_{nod}}$ or a self-adjoint elliptic differential operator of second order like the Laplacian, $H_j u_j= \mathrm{div}(a \nabla u_j)$, $a \in C^\infty(\overline{\Omega}_j)$, with Neumann boundary conditions. The network coupling $T_{ij}$ consists of disjoint sinks and sources, as in \Cref{FigureMetaplex}. The amplitude is chosen according to the $d$-path Laplacian \eqref{kpathmel}, \eqref{kpathlap}, with hopping between $\Omega_v$ and $\Omega_w$ proportional to $\mathrm{dist}(v,w)^{-s_{net}}$ (long range coupling), respectively $2^{-s_{net}\mathrm{dist}(v,w)}$ (short range coupling). 


When we consider a metaplex we are dealing with the endo-dynamics
occurring inside the nodes and the exo-dynamics occurring between them. Inside each node $\Omega_j$ the solution $e^{t H_j} u_{j,0}$ to the diffusion equation for $H_j$ with initial condition $u(t=0) = u_{j,0}$ is an integral operator $\left(e^{t H_j} u_0\right)(t,\mathbf{x}) = \int_\Omega K_{H_j}(t,\mathbf{x},\mathbf{y}) u_0(\mathbf{y})\ d\mathbf{y}$. The  integral kernel $K_{H_j}$ describes the evolution of a Dirac point mass in $\mathbf{y}$ at time $t=0$. For ordinary diffusion away from the boundary $\partial \Omega_j$ it satisfies the Gaussian estimate 
\begin{equation}\label{gaussest}
|K_{H_j}(t,\mathbf{x},\mathbf{y})| \leq Ct^{-n/2}e^{-C \frac{|\mathbf{x}-\mathbf{y}|^2}{t}}
\end{equation}
for some $C>0$, analogous to the explicit solution formula for the diffusion equation on $\mathbb{R}^n$  \cite{ouhabaz2009analysis}. Similar bounds are known for ordinary diffusion on metric measure spaces $\Omega_j$ \cite{grig}. Fractional diffusion, on the other hand, exhibits slow algebraic decay in the form of a Poisson estimate
\begin{equation}\label{poissonest}
|K_{H_j}(t,\mathbf{x},\mathbf{y})| \leq Ct \left(|\mathbf{x}-\mathbf{y}| + t^{1/(2s_{nod})}\right)^{-n-2s_{nod}}
\end{equation}
away from  $\partial \Omega_j$  \cite{Gimperlein2014}. In a convex domain the estimate is sharp for short times, and it allows us to estimate the diffusion between nodes across the network.

Consider the linear network in \Cref{FigureMetaplex}, with an initially uniform density localized in $\Omega_1$. From \Cref{sourcecoupling} we calculate the change of the total density in node $j$: 
\begin{align}
\partial_t\Big|_{t=0} \int_{\Omega_j} u_j(t,\mathbf{x})\ d\mathbf{x}=  - \frac{\delta_{j1}}{|\Omega_1|} \sum_{i}\int_{\Omega_j}\alpha_{ij}(\mathbf{x})d\mathbf{x} +  \frac{1}{|\Omega_1|} \int_{\Omega_j} \alpha_{j1}(\psi_{j1}^{-1}(\mathbf{x}) \mathrm{det}(\nabla \psi_{j1}^{-1}) \  d\mathbf{x} \ .\end{align}
So, initially particles hop from node $1$ to node $j$ according to the transition probabilities $\alpha_{ij}$ of the network, from $\mathbf{x}$ to $\psi_{j1}(\mathbf{x})$. From Duhamel's formula for the solution of the inhomogeneous diffusion equation, they then evolve inside $\Omega_j$ according to $K_{H_j}$ before jumping back to node 1, or further to a different node $k$. Both processes only enter into the next, quadratic term of the Taylor expansion in $t$. These formal arguments are made rigorous in terms of an asymptotic expansion for $t \to 0$ of the heat kernel $K_\mathcal{D}$ for the network of interacting domains, see e.g., \cite{dewitt}.

Compared to nodes without internal structure, hopping to node $k$ is reduced by the rate $K_{H_j}(\tau,\psi_{j1}(\mathbf{x}),\mathbf{y})$ of diffusing from $\psi_{j1}(\mathbf{x})$ to a point $\mathbf{y}$ in the region of the sink to node $k$ within time $\tau$. The internal diffusion thus always slows down the network dynamics if sources and sinks are disjoint. 

The exit time $\tau^j_{source,\ sink}$ taken to get from source to sink is well studied both for normal and fractional diffusion \cite{bertoin}. It satisfies a (fractional) Fokker-Planck equation and satisfies the estimate \eqref{gaussest}, respectively \eqref{poissonest}, for short times. For long times, $\tau^j_{source,\ sink}$ decays exponentially fast in time. 

For short times, the dependence on the internal diffusion $H_j$, the geometry of $\Omega_j$, and the location of sinks and sources are well described by \eqref{gaussest} and \Cref{poissonest}. For sources and sinks which are far apart, in the case of normal diffusion \Cref{gaussest} shows a faster than exponential suppression  of the transition rate due to the internal structure. Fractional diffusion suppresses the transition rate algebraically in the distance between sink and source, according to \Cref{poissonest}. The smaller $s_{nod}$, the faster the equilibration and the smaller the suppression in $\Omega_j$. 

Similarly if $T_{ij}$ is nonzero only for $|i-j|\leq 1$, hopping to next $j$-th nearest neighbors is suppressed $j$ times by the internal diffusion. Starting from an initial density localized in node $1$, the total density in node $j$ will be exponentially small, $\int_{\Omega_j} u_j(t,\mathbf{x})\ d\mathbf{x} \leq C e^{-C|j-1|}$ for some $C = C(t)>0$ depending on the endo-structure.

\section{Numerical analysis of diffusion in metaplexes}\label{sec: numerics}

\begin{figure}[tbhp]
  \centering
  \subfloat[]{\label{fig: density profiles a}\includegraphics[width = 0.5\textwidth]{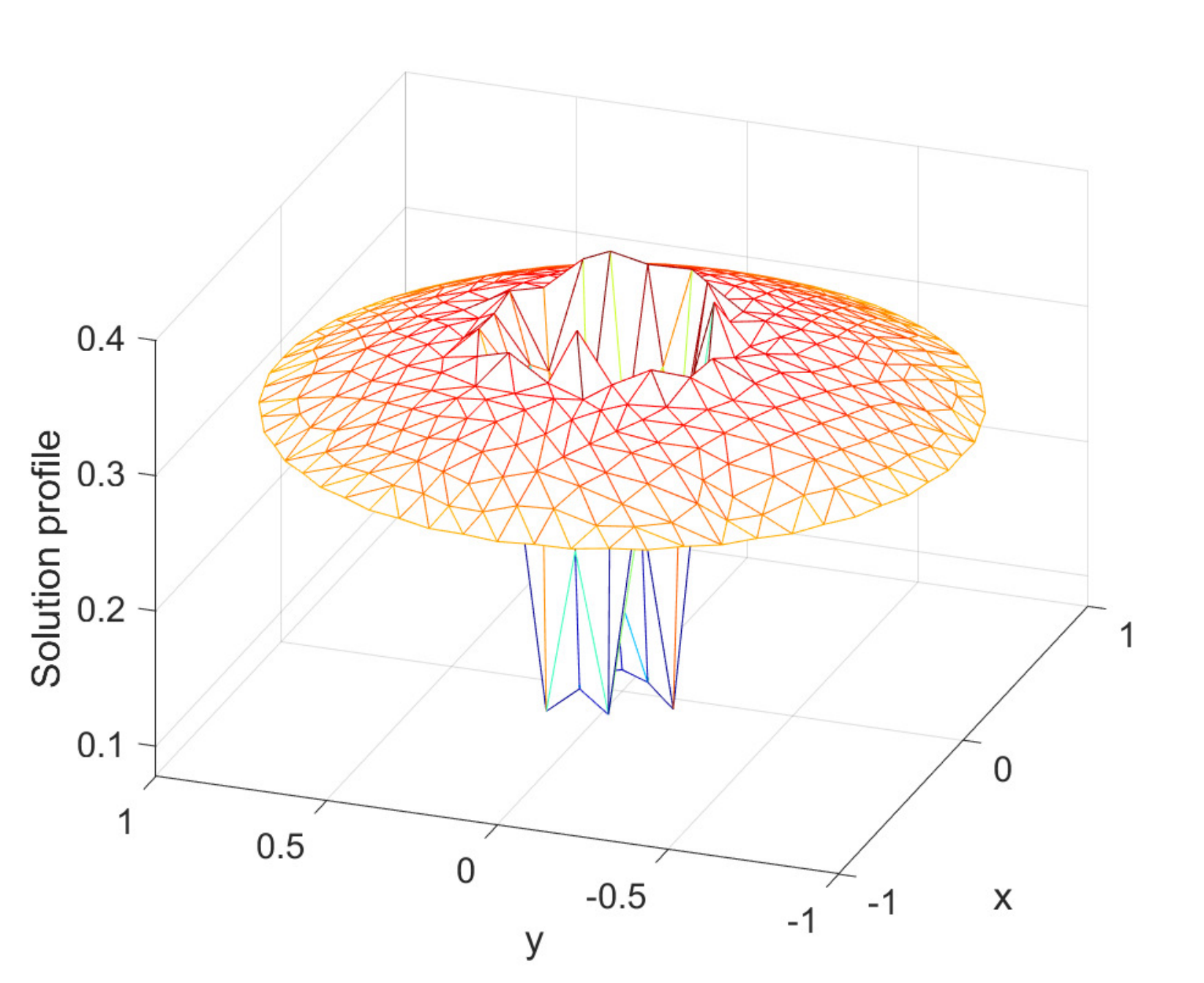}}
  \subfloat[]{\label{fig: density profiles b}\includegraphics[width=0.5\textwidth]{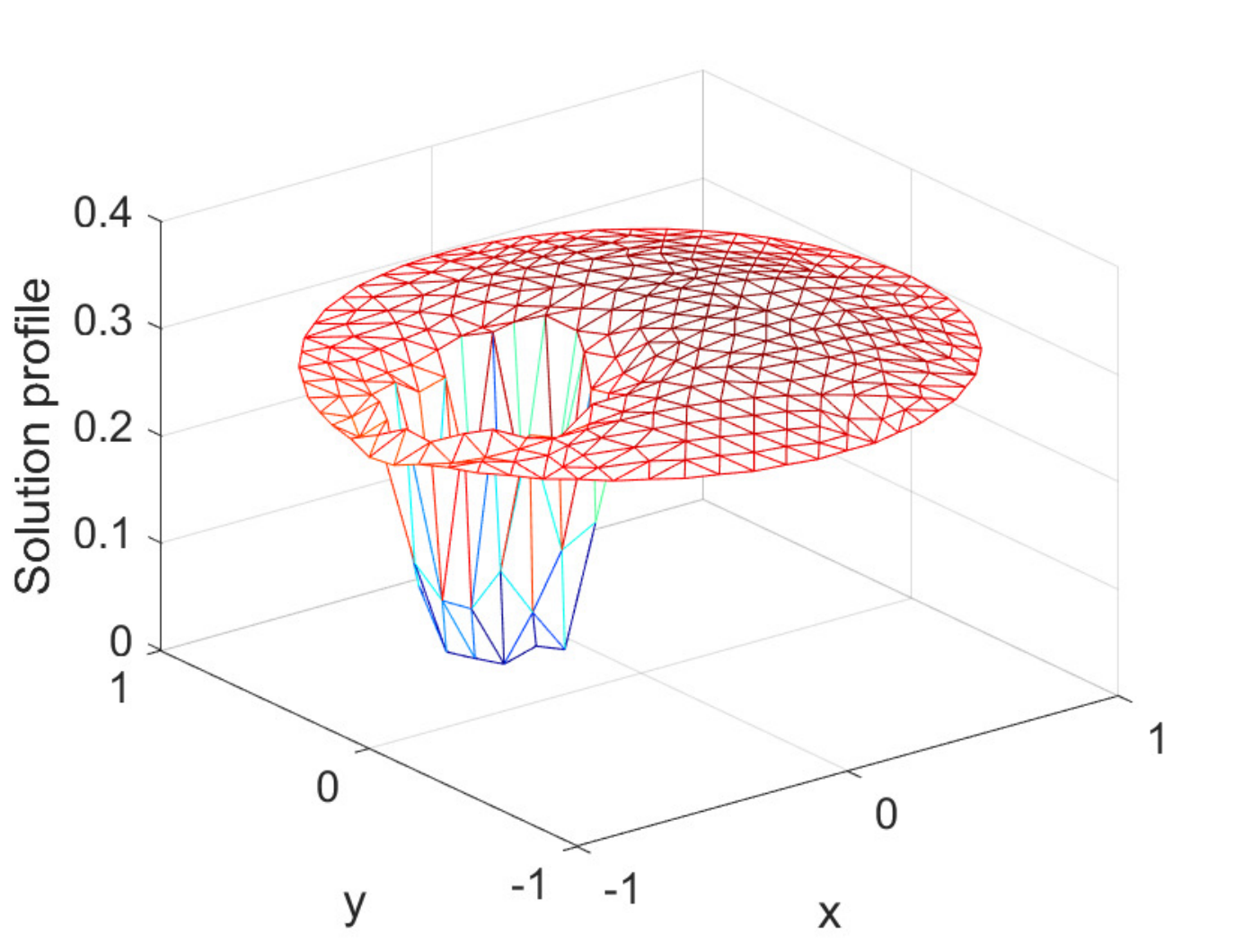}}  
  \caption{Density profile for different coupling points between nodes.}\label{fig:density profiles coupling}
\end{figure}

This section  investigates the influence of the internal structure of the nodes on the global metaplex dynamics, using numerical experiments for the toy example {of a linear metaplex}.  For the localized coupling $T_{ij}$ in a given area (\cref{fig:density profiles coupling}) we show the rich dependence on the internal structure in a metaplex {such as the geometry of the nodes and the nature of the coupling, and put  the results, which we obtained in certain limits, into a wider perspective. For example, a superdiffusive process in the nodes may \emph{slow down}  equilibration in the metaplex, see Experiment 1. On the other hand, the network diffusion dominates the qualitative global behavior: In \Cref{superdiffsection} we show that network superdiffusion due to long range hopping survives independently of the internal structure.
This section illustrates the rich phenomena which arise from the interplay of the exo- and endo-structures of the metaplex. See Supplementary Information for numerical and analytical results for the case $T_{ij}=\textnormal{Id}$. The conclusions are not restricted to the linear metaplex, but will be studied in the subsequent  section in some  real-world examples.}

\subsection{Set up of numerical experiments}

Here we analyze a simple metaplex consisting of $51$ identical
circular domains $\Omega=\Omega_{j}\subset\mathbb{R}^{2}$ connected
in the form of a linear chain, i.e., a path graph. The nodes are labelled
in consecutive order starting by one. Starting from a uniform distribution
in node 1, we study the evolution of the density in the nodes depending
on the network and the internal diffusion processes, the size of the
nodes and the strength and nature of the coupling between the nodes.

\begin{figure}[tbhp]
  \centering
  \subfloat[]{\label{fig: mesh1}\includegraphics[width = 0.3\textwidth]{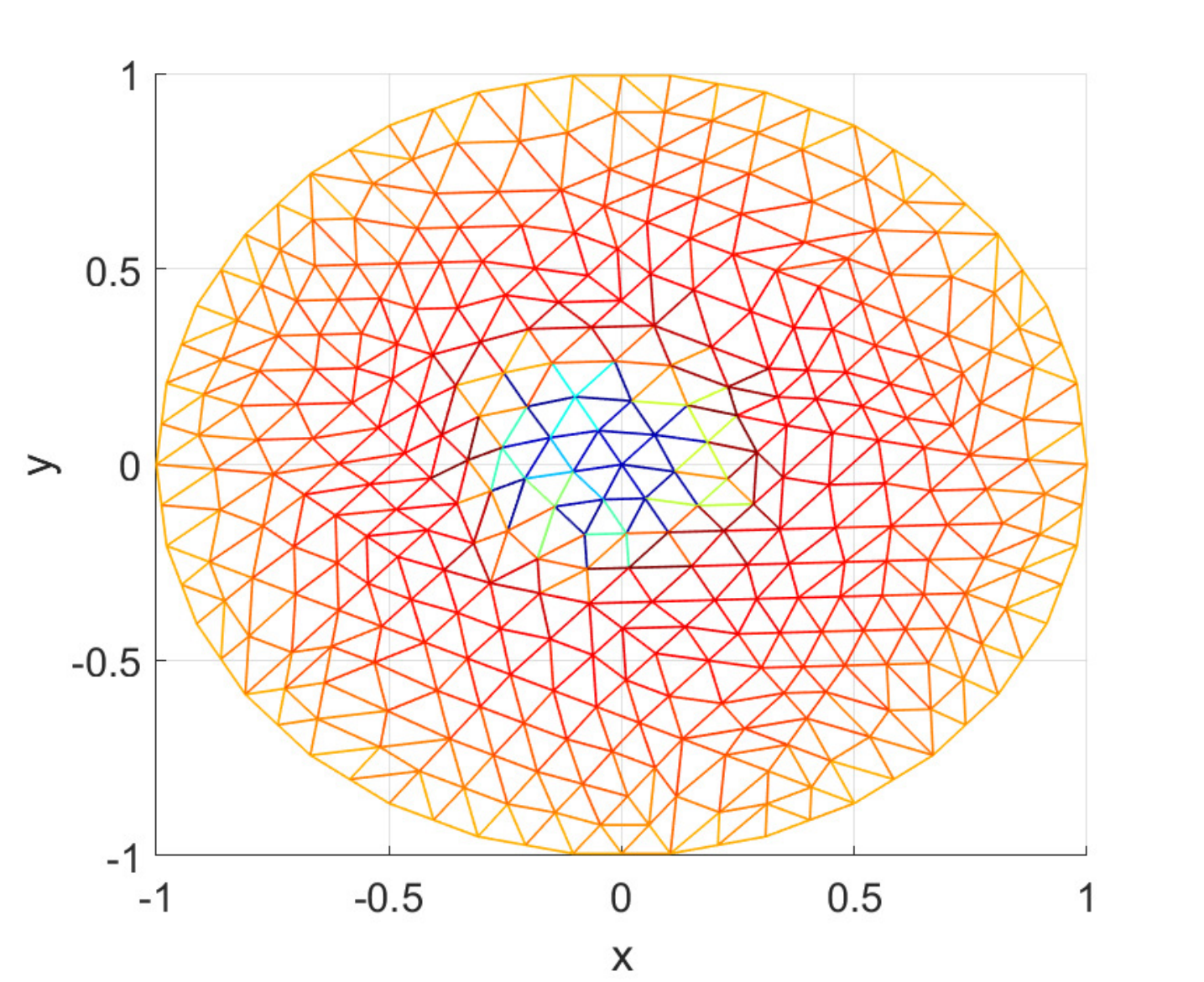}}
  \subfloat[]{\label{fig: mesh2}\includegraphics[width = 0.3\textwidth]{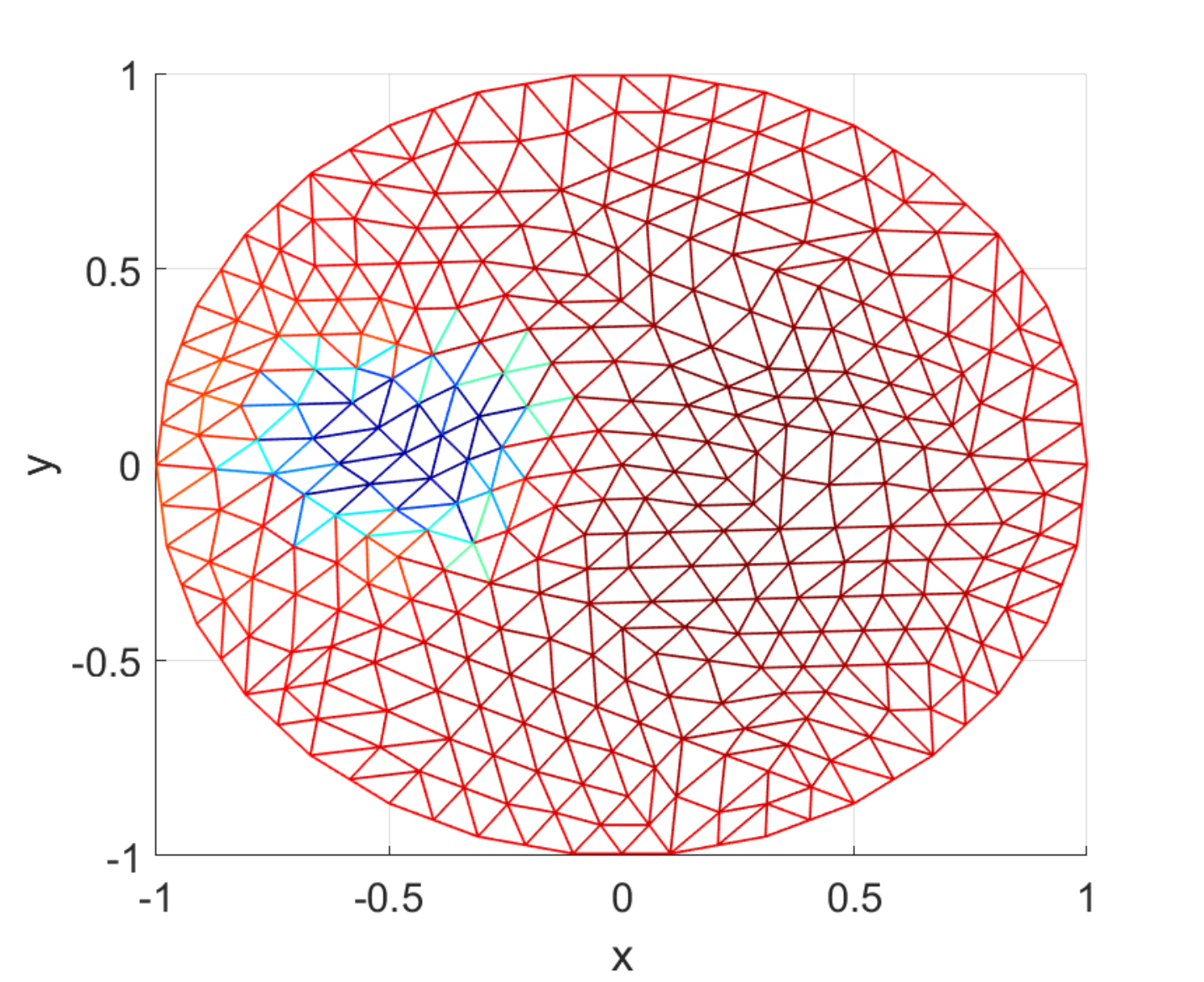}}  \subfloat[]{\label{fig: mesh3}\includegraphics[width = 0.3\textwidth]{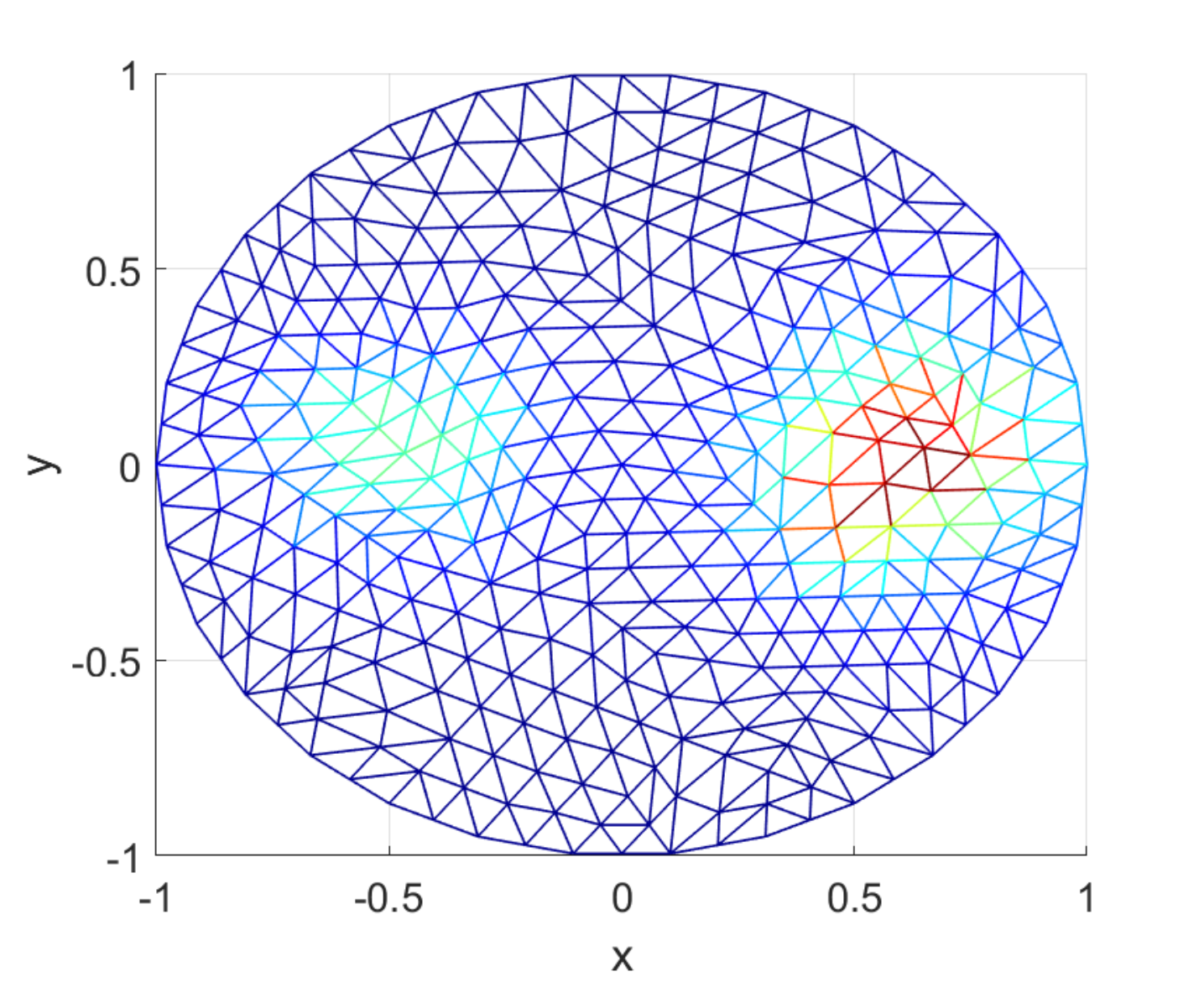}}
  \caption{Mesh inside the nodes for the different coupling areas in experiment 1 (a) and experiment 2 (b, c).}\label{fig: mesh}
\end{figure}

Two different sizes of nodes are considered, $\Omega_{s}=B(0,1)$,
$\Omega_{b}=B(0,100)$. These different sizes will also be used when
we move to the analysis of real-world metaplexes. The diffusion equation \eqref{sourcecoupling} inside each node with L\'{e}vy exponent $s_{nod}$  is approximated by finite elements in space on a quasi-uniform spatial mesh with 347 degrees of freedom. See \cref{fig: mesh} for a plot of the mesh and \cite{m3as} for the numerical approximation of the fractional Laplacian. For the time discretization we use a backward Euler method in time with a sufficiently small, fixed time step  $dt=0.01$. 

Unless stated otherwise, the network coupling between nodes $(i,j)$ is taken to be of short range, according to {the Laplace-transformed $d$-path Laplacian \eqref{kpathlap}. The coupling strength between nodes $(i,j)$ is thus proportional to $2^{-s_{net}|i-j|}$ for different exponents $s_{net}$. }

\subsection{Experiment 1: Central coupling region}\label{subsec: experiment1}
For different values of the network exponent $s_{net} = 0.4, 0.8$, we compare close to normal diffusion (L\'{e}vy exponent $s_{nod} =0.8$) to superdiffusion ($s_{nod} =0.2$) inside a metaplex of small nodes $\Omega_s$.
In this experiment we consider that the nodes of the metaplex are
connected by means of a central region of the node, as depicted in \cref{fig: density profiles a}, which acts at
the same time as a sinks and as a source. See Supplementary Material (Note 2) for a more detailed description of sink and sources.  The coupling strength is fixed to be $\alpha=10$.

\begin{figure}[!ht]
    \centering
    \includegraphics[scale=0.4]{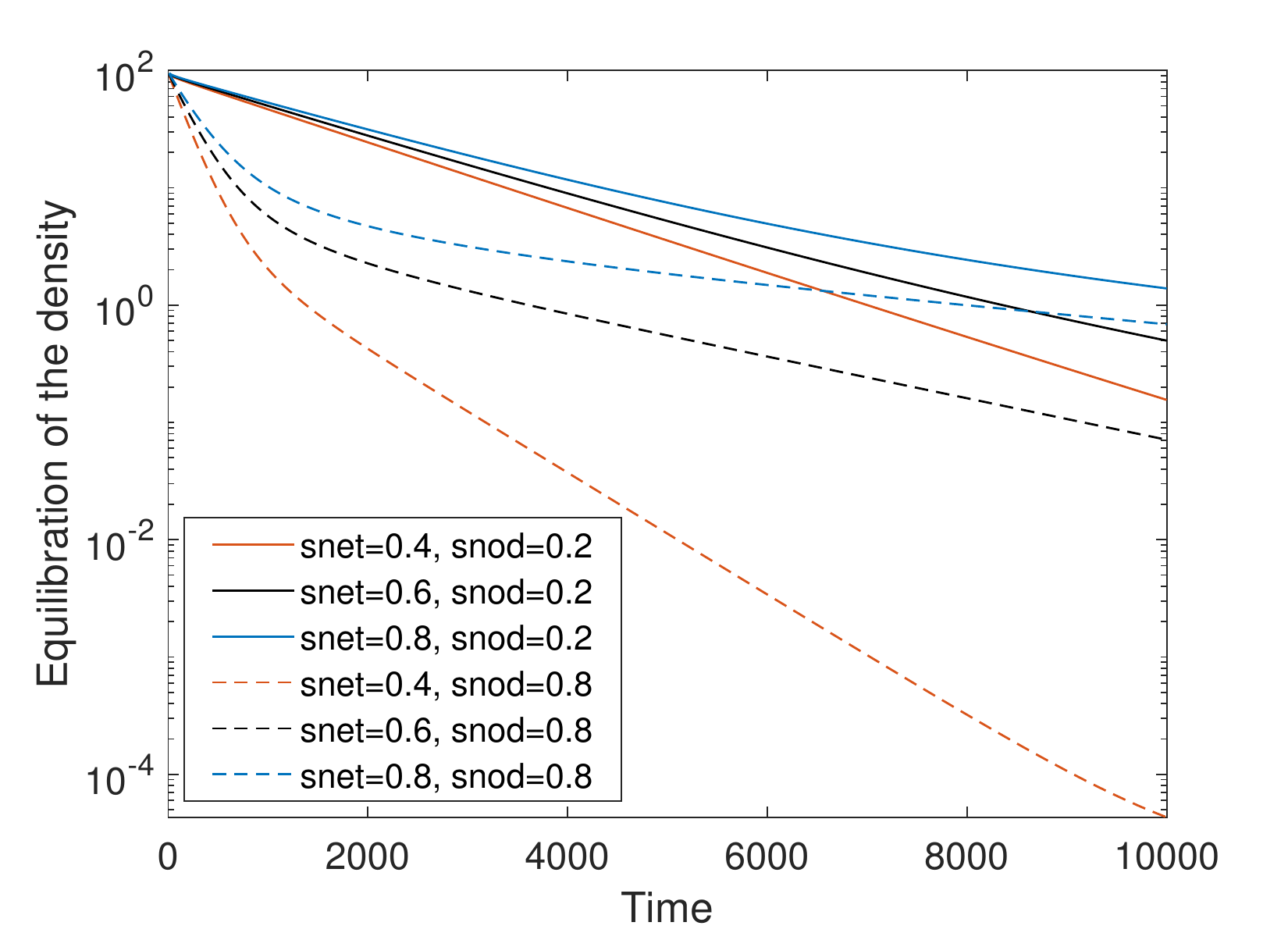}
    \caption{Density equilibration in node 1 for central coupling region.}
    \label{fig: densities one coupling point}
\end{figure}

\cref{fig: densities one coupling point} shows the time at which the density in node 1 reaches the equilibrium given by $|\int_{\Omega_1} u_1(t,\mathbf{x})d\mathbf{x}-\frac{1}{N}\int_{\Omega_1}u_{0}(\mathbf{x})d\mathbf{x}|$, where $u(t,\mathbf{x})$ is the density in the node at a given time $t$ and $u_0(\mathbf{x})$ is the initial density in the metaplex. The figure illustrates the strong effect of the dynamics inside the nodes on the diffusion process, even though the networks dynamics dominates. We see that superdiffusion inside each node slows down the diffusion in the network: Due to the nature of the L\'{e}vy process, particles quickly diffuse far away from the sink. As they take time to return and hop to the next node, diffusion between nodes in the whole network is slowed down. In \Cref{subsec: experiment2} we discuss how the size of the nodes affects this behavior.

The spatial localization of the sink and sources decreases the total strength of the coupling, resulting in a slower equilibration of the densities in the node compared to a uniform coupling discussed in the Supplementary Material (Note 3). Changing the strength of a sufficiently strong localized coupling does not affect this behavior much, as the equilibration time saturates: independent of the coupling strength, particles which are located away from the sink cannot hop to a neighboring node.
\begin{figure}[!ht]
    \centering
    \includegraphics[scale=0.4]{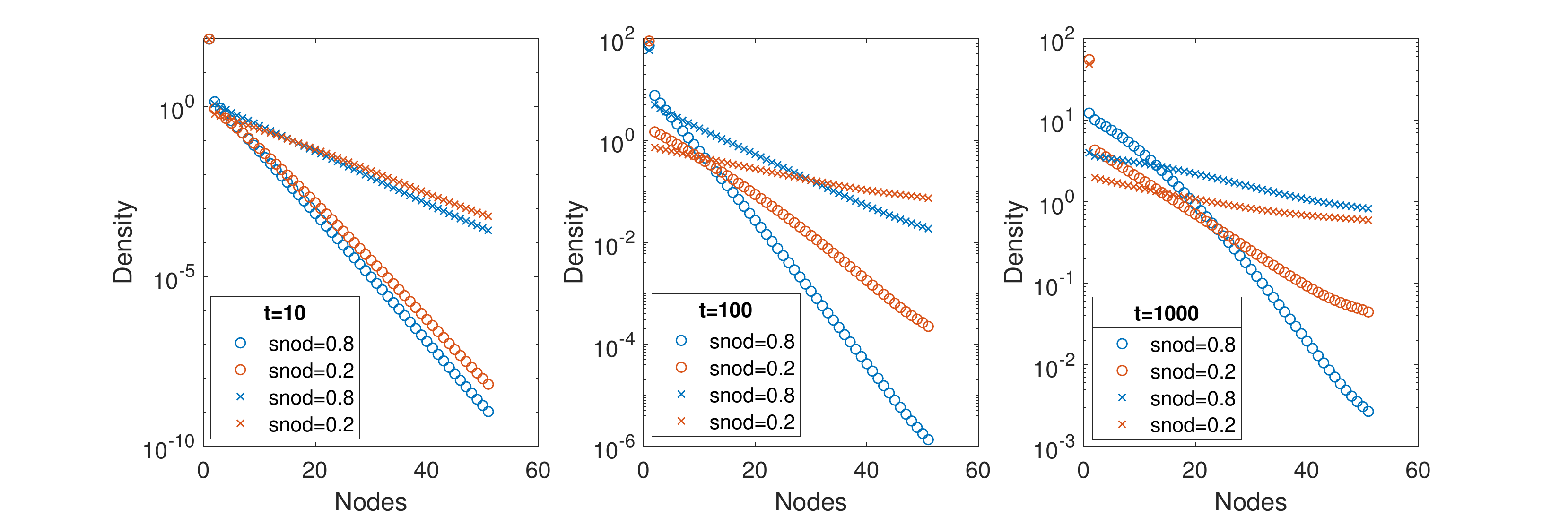}
    \caption{Density distribution for central coupling region, $s_{net}=0.8$  ($\circ$), resp.~$0.4$ ($\times$).}
    \label{fig: density vs nodes}
\end{figure}

In \cref{fig: density vs nodes} we show the evolution in time of the total density $\int_{\Omega_j} u_j(t,\mathbf{x}) d\mathbf{x}$ in each node $\Omega_j$. While superdiffusion inside the nodes has only limited effect on the equilibration of the densities in the network, it speeds up hopping to distant nodes. From \cref{fig: density vs nodes} we observe that, when the network dynamics is dominated by superdiffusion ($s_{net}=0.4$), only for short times the superdiffusion inside the nodes (red crosses) overtakes normal diffusion, while for longer times the equilibrium is attained faster for $s_{nod}=0.8$.

\subsection{Experiment 2: Distant coupling regions}\label{subsec: experiment2}

To study the effect of the {geometry of the coupling and the node} on the global dynamical process, we {replace the central coupling from  \Cref{subsec: experiment1}} by a prototypical metaplex coupling as depicted in \cref{FigureMetaplex}: For the odd nodes the coupling region is located as in \cref{fig: density profiles b}, while for even nodes the coupling region is on the opposite side. Similar to the central coupling, the coupling regions act as sink and as sources at the same time.  The coupling areas in both domains are equal, therefore, in the case of $\Omega_b$, the areas are more localized and distant. 

For the network dynamics, we consider the short range coupling from Experiment 1 given by {the Laplace-transformed $d$-path Laplacian \eqref{kpathlap}, with coupling between nodes $(i,j)$ proportional to} $2^{-s_{net}|i-j|}$, $s_{net} = 0.4, 0.8$. The coupling strength is fixed to be $\alpha=10$ for $\Omega_s$ and $\alpha=100$ for $\Omega_b$. In \Cref{superdiffsection} we compare the resulting dynamics  to a long range coupling proportional to $|i-j|^{-s_{net}}$, with the Mellin-transformed $d$-path Laplacian \eqref{kpathmel}.

\begin{figure}[!ht]
\centering
 \includegraphics[scale=0.4]{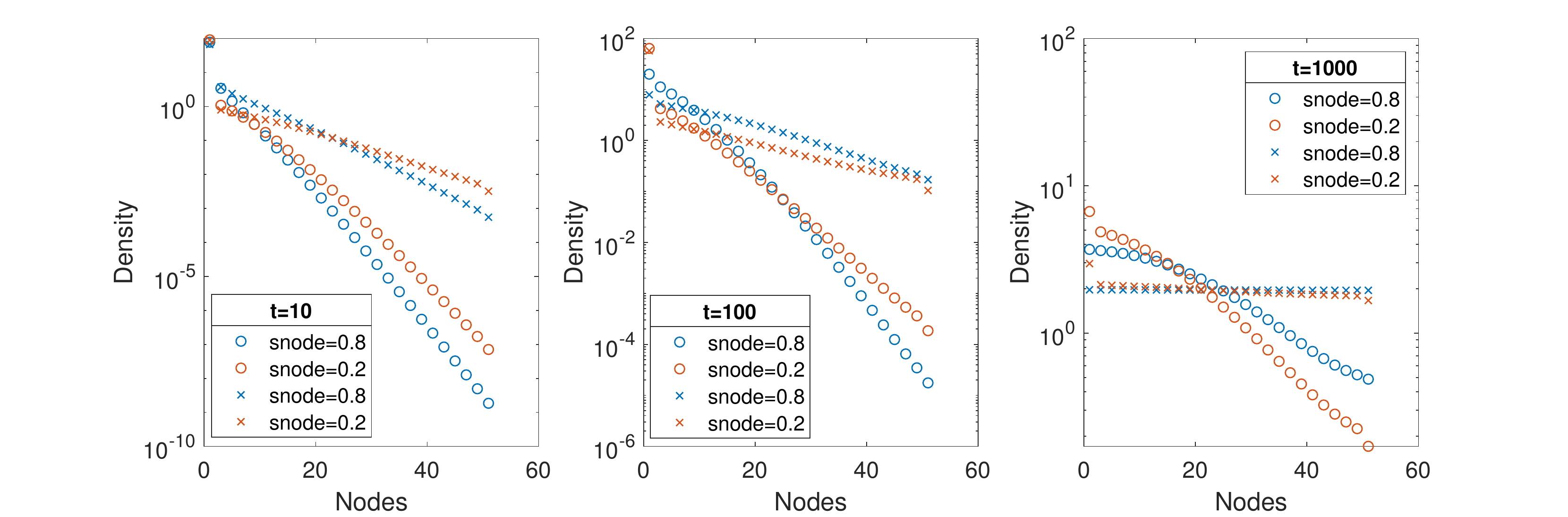}
    \caption{Density distribution for small nodes $\Omega_s$, $s_{net}=0.8$  ($\circ$), resp.~$0.4$ ($\times$).}
    \label{fig: density vs nodes different coupling}
\end{figure}

\begin{figure}[!ht]
\centering
   \includegraphics[scale=0.4]{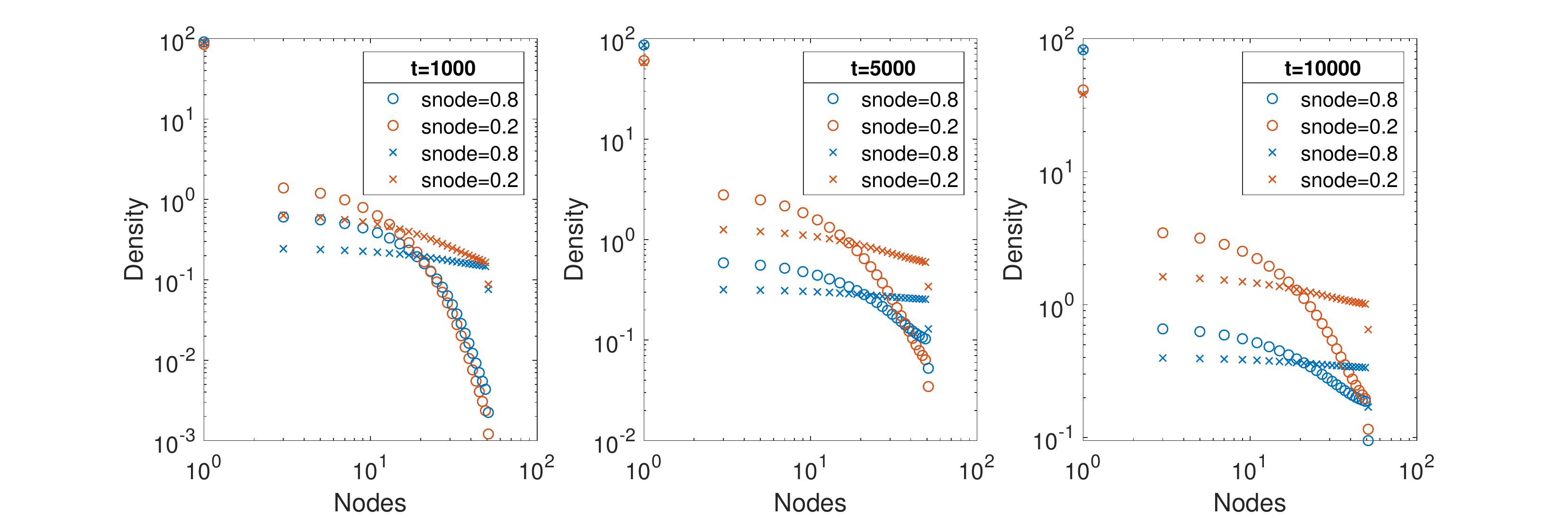}
\caption{Density distribution for big nodes $\Omega_b$, $s_{net}=0.8$  ($\circ$), resp.~$0.4$ ($\times$).}\label{fig: bignodes}\label{fig: big node density-time}
\end{figure}

In the current experiment, the distance between the sinks and sources
also leads to a delay; particles appear in the node and have to diffuse
to find the sink to hop to another node. Coupled with the exo-dynamics,
the delay leads to small density oscillations between neighboring
nodes. For clearer illustration in \cref{fig: density vs nodes different coupling,fig: bignodes,fig: normal diffusion Mellin,fig: mellin superdiffusion} we only consider the odd nodes of the metaplex.

\begin{figure}[tbhp]
  \centering
  \subfloat[]{\label{fig: density different sizes a}\includegraphics[width = 0.45\textwidth]{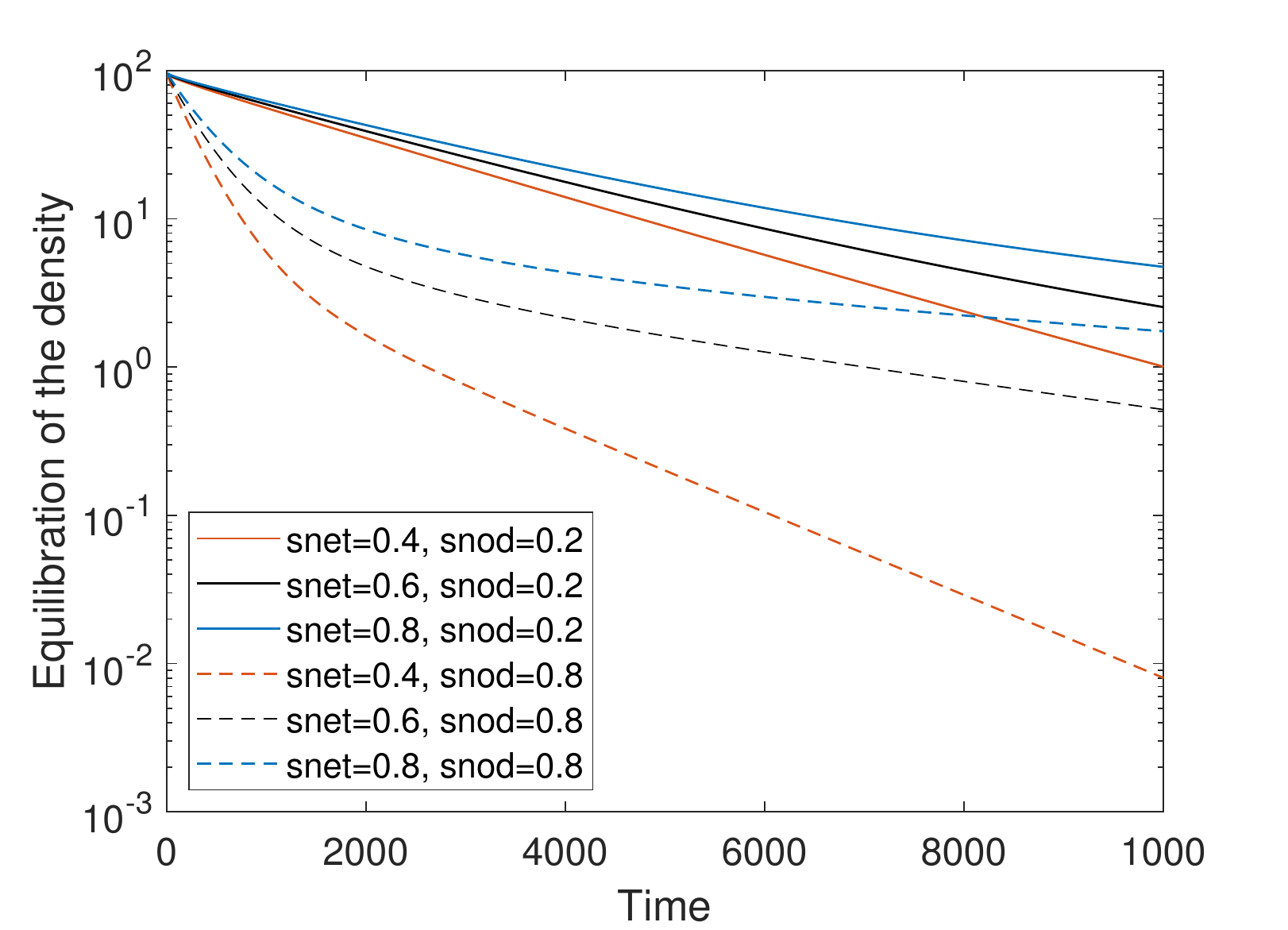}}
  \subfloat[]{\label{fig: density different sizes b}\includegraphics[width=0.45\textwidth]{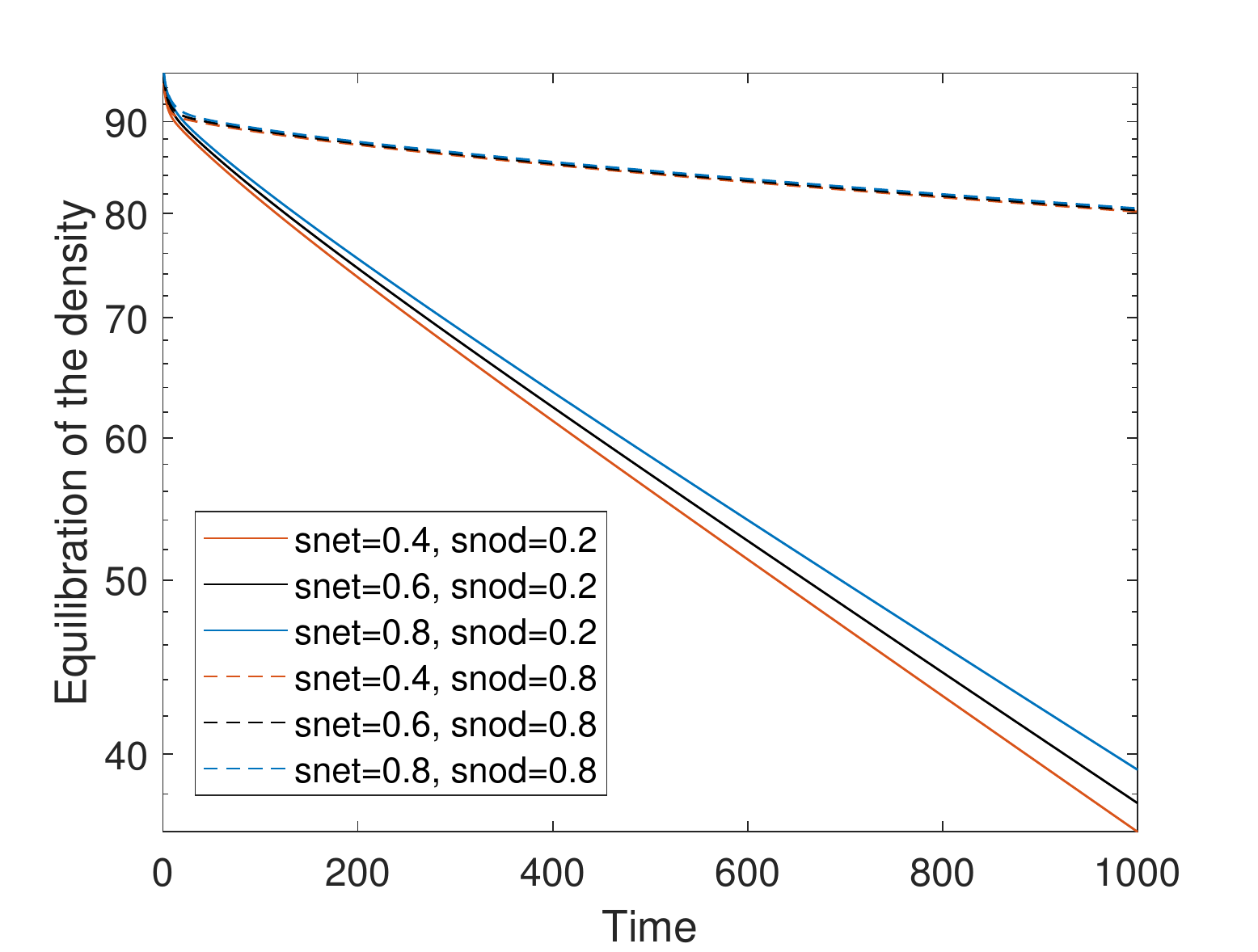}}  
  \caption{Density equilibration in node 1 for disjoint sinks and sources (a) $\Omega_s$ (b) $\Omega_b$.}\label{fig: density different sizes}
\end{figure}


\cref{fig: density different sizes a} shows the deviation of the density in node 1 from equilibrium, corresponding to \cref{fig: densities one coupling point} in Experiment 1.  Similar to Experiment 1, superdiffusion inside the small nodes $\Omega_s$ slows down equilibration. However, the evolution in time of the total density $\int_{\Omega_j} u_j(t,\mathbf{x}) d\mathbf{x}$ in  nodes $\Omega_i$ is depicted in \cref{fig: density vs nodes different coupling}, at times $t=10,100,1000$. One observes that superdiffusion in the nodes allows particles to reach distant nodes more efficiently than approximately normal diffusion $s_{nod}=0.8$. This confirms the interpretation in \Cref{sec: operators}. The case $s_{net}=0.4$ in \cref{fig: density vs nodes different coupling} exhibits similar dynamics as in Experiment 1, where $s_{nod}=0.8$ equilibrates the metaplex's density faster on long time scales.

We  conclude that particles undergoing approximately normal diffusion ($s_{nod}=0.8$) are slower, but more precise. Superdiffusing  particles ($s_{nod}=0.2$)  explore the metaplex faster, but take more time to equilibrate the density across the whole metaplex. Similar observations have been made   in \cite{matthaus2009coli}, where the targeting efficiency of  \emph{E. coli} bacteria was studied in simulations. They observed that bacteria with higher motility, following a superdiffusion process, found targets faster, but were also at risk of moving rapidly away from the target due to the nature of their L\'{e}vy walk. Individuals with lower motility were slower but more precise.

\begin{figure}[tbhp]
  \centering
  \subfloat[]{\label{fig: mellin densities a}\includegraphics[width = 0.45\textwidth]{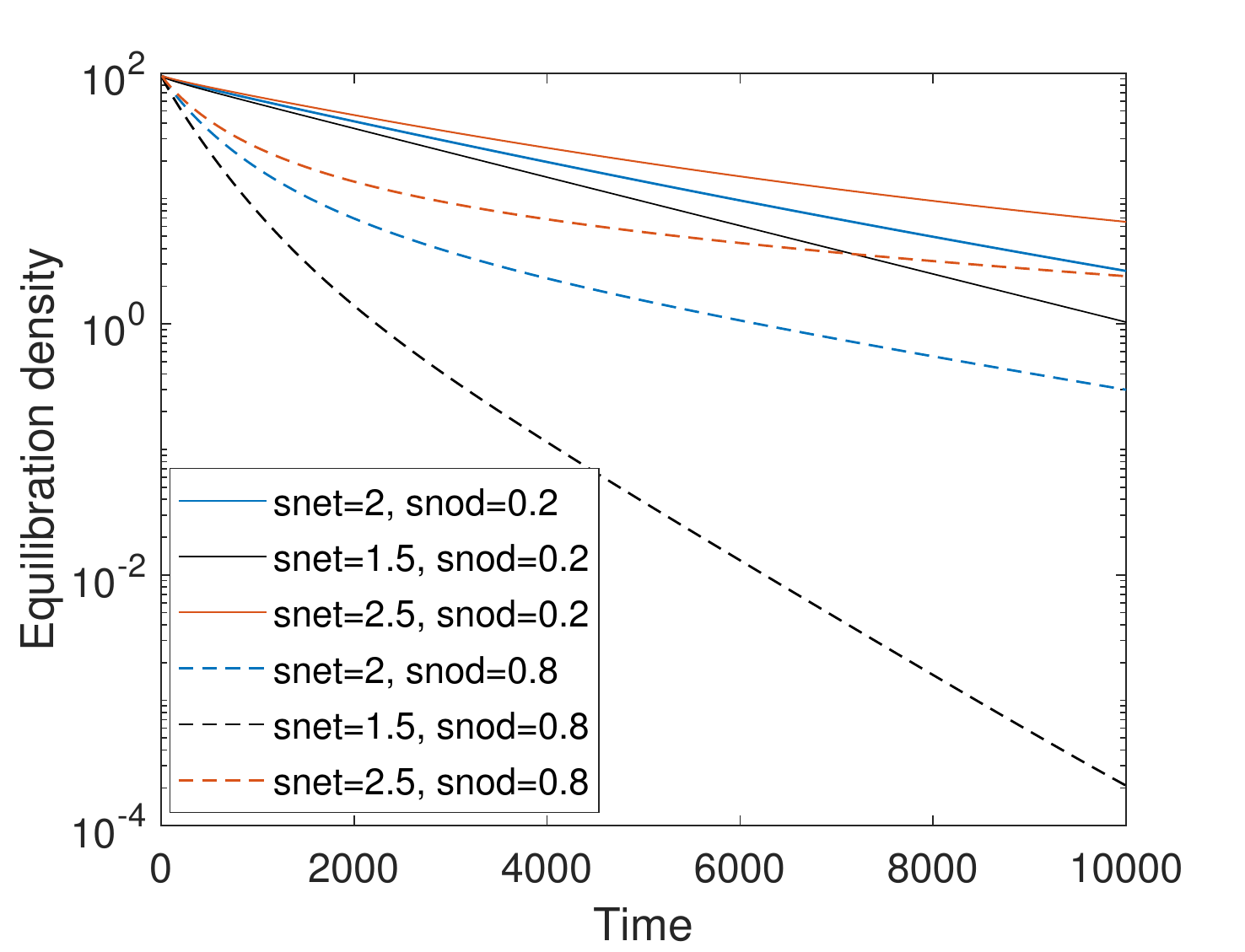}}
  \subfloat[]{\label{fig: mellin densities b}\includegraphics[width=0.45\textwidth]{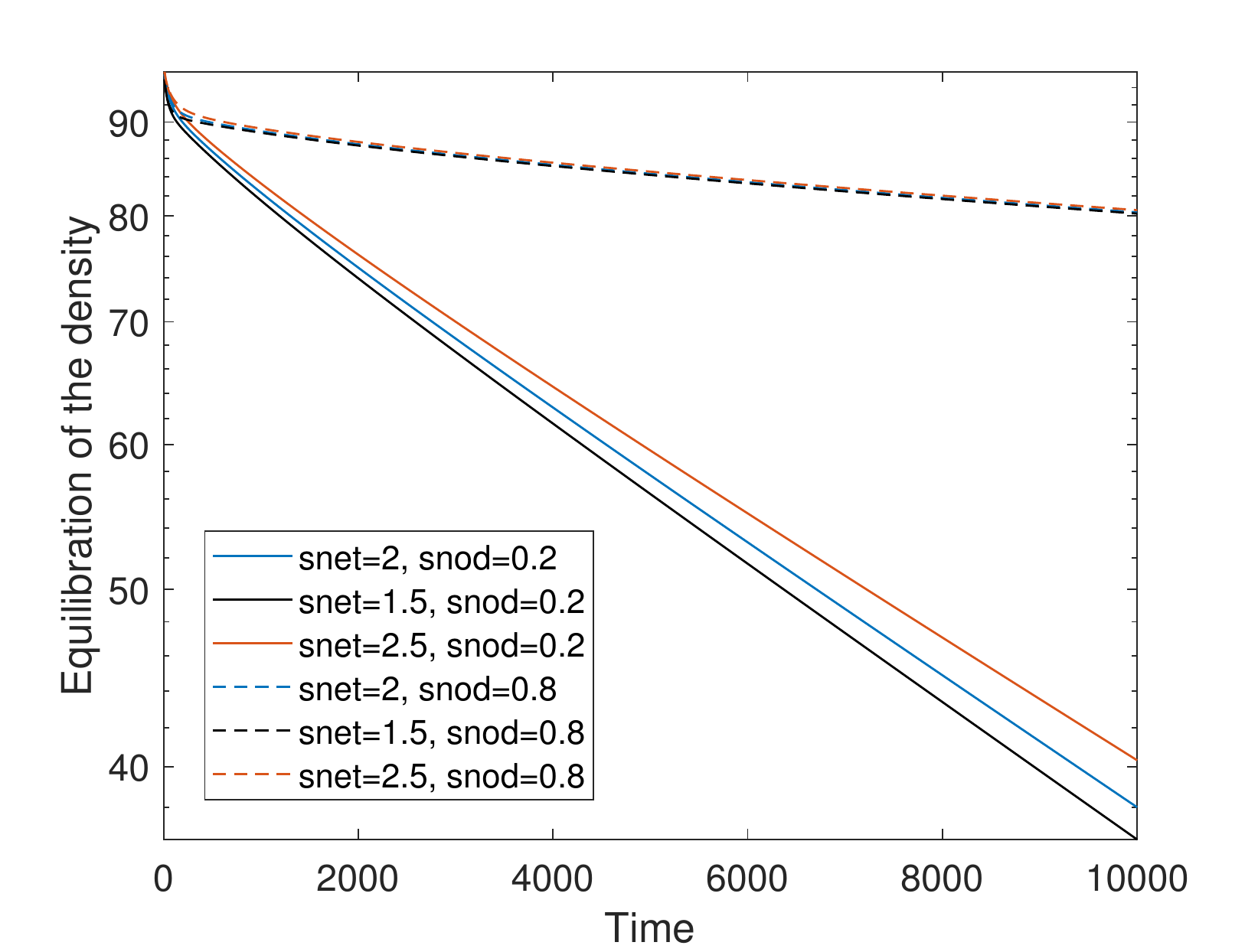}}  
  \caption{Density equilibration in node 1 for the Mellin-transformed $d$-path Laplacian for $\Omega_s$ (a) and $\Omega_b$ (b).}\label{fig: mellin densities}
\end{figure}

The following figures compare these conclusions to those obtained for a network of big nodes $\Omega_b$. 

In \cref{fig: density different sizes b} we observe that superdiffusion inside big nodes speeds up the equilibration of the densities considerably: For $s_{nod}=0.2$ particles require a much smaller time to reach their distant target, the sink, than for $s_{nod}=0.8$. This is in agreement with the discussion in \Cref{sec: variation of eig}.

In \cref{fig: bignodes} we plot the density as a function of the node, in a log-log plot at times $t=1000, 5000, 10000$. Note that because of the large distance between sources and sinks for nodes $\Omega_b$, the time scale to approach equilibrium increases significantly. We observe that superdiffusion inside the nodes accelerates  the equilibration over the whole metaplex in all cases, unlike for the small nodes $\Omega_s$ (\cref{fig: density vs nodes different coupling}). For $s_{net}=0.4$ in \cref{fig: bignodes} we observe the accelerated diffusion clearly, with a density distribution which is far from a Gaussian parabola in the log-log plot.

\subsubsection{Metaplex superdiffusion}\label{superdiffsection}

From the discussion of Experiments 1 and 2 so far, we conclude that for the short ranged network diffusion \eqref{kpathlap}, proportional to $2^{-s_{net}|i-j|}$, superdiffusion in the nodes can accelerate the equilibration in the metaplex, but it cannot lead to global superdiffusion. This is in line with the discussion in \Cref{sec: analysis of diffusion in metaplexes}. We now consider a long ranged network  coupling proportional to $|i-j|^{-s_{net}}$, given by the Mellin-transformed $d$-path Laplacian \eqref{kpathmel} for $s_{net}=1.5,\ 2,\ 2.5$ and $4$. 

In the case of this linear metaplex we count with the advantage that
previous analytic results exists for the diffusion on its exo-skeleton
using the $d$-path Laplacians \cite{estrada2012path}. In this case it was proved analytically that the use of the
Mellin transform may produce superdiffusive regime in an infinite
linear chain of nodes for $s_{net} \in (1,3)$.

\begin{figure}[tbhp]
    \centering
    \includegraphics[scale=0.4]{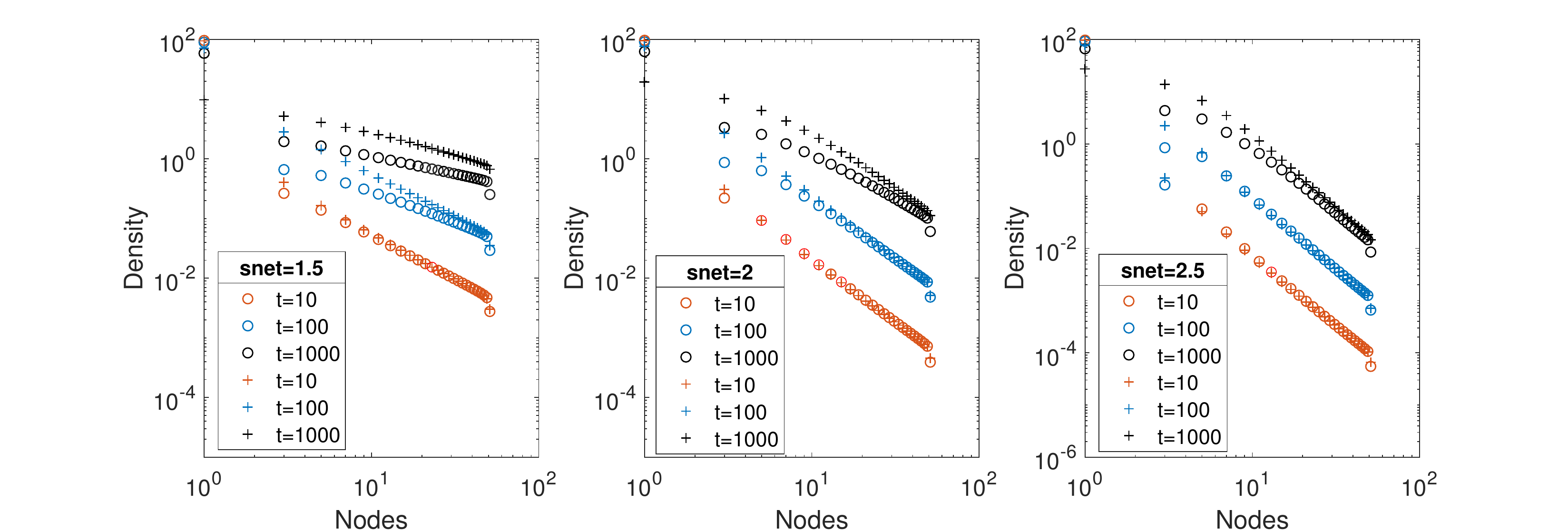}
    \includegraphics[scale=0.4]{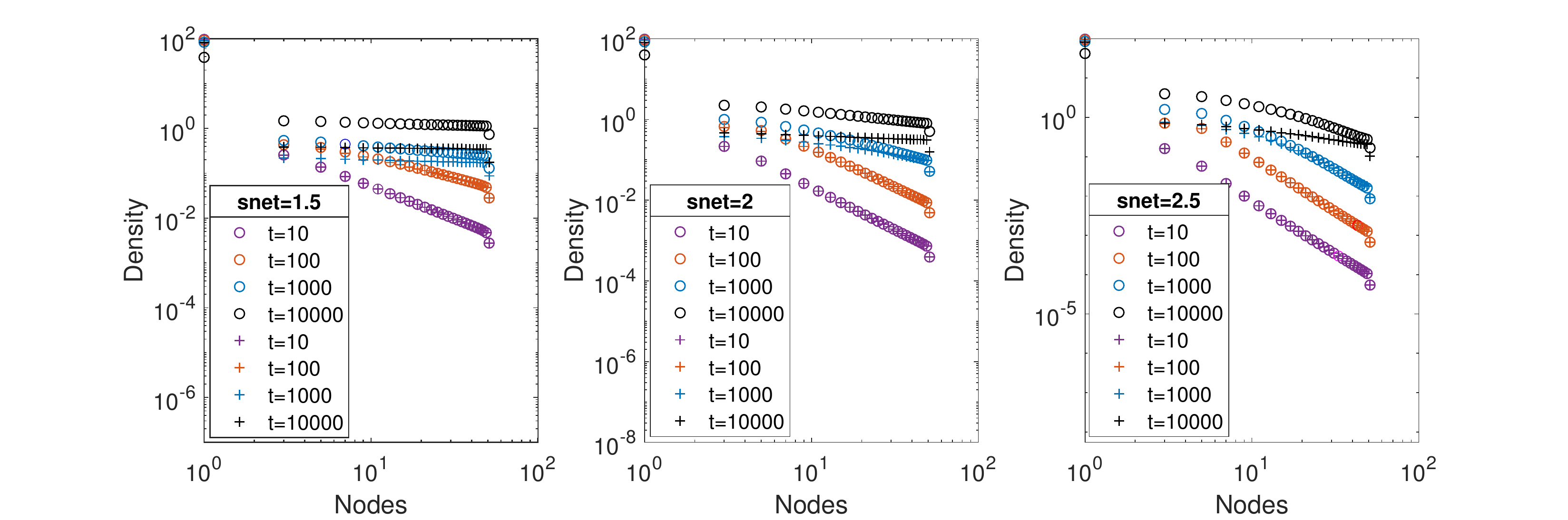}
    
    \caption{Density distribution for long ranged network coupling, $s_{nod}=0.2$ ($\circ$) and $s_{nod}=0.8$ ($+$). Top panel: $\Omega_s$. Bottom panel: $\Omega_b$ }
    \label{fig: mellin superdiffusion}
\end{figure}

As in \cref{fig: bignodes}, \cref{fig: mellin superdiffusion} plots the density at times $t=10,100,1000,10000$ as a function of the node of the network in a log-log plot. The linear decay of the density and the peaked behaviour at node 1 indicate superdiffusion irrespective of the internal dynamics. Because of the strong network diffusion, for the big nodes $\Omega_b$ the structure of the nodes proves irrelevant to the metaplex dynamics.

\begin{figure}[!ht]
\centering
\includegraphics[scale=0.4]{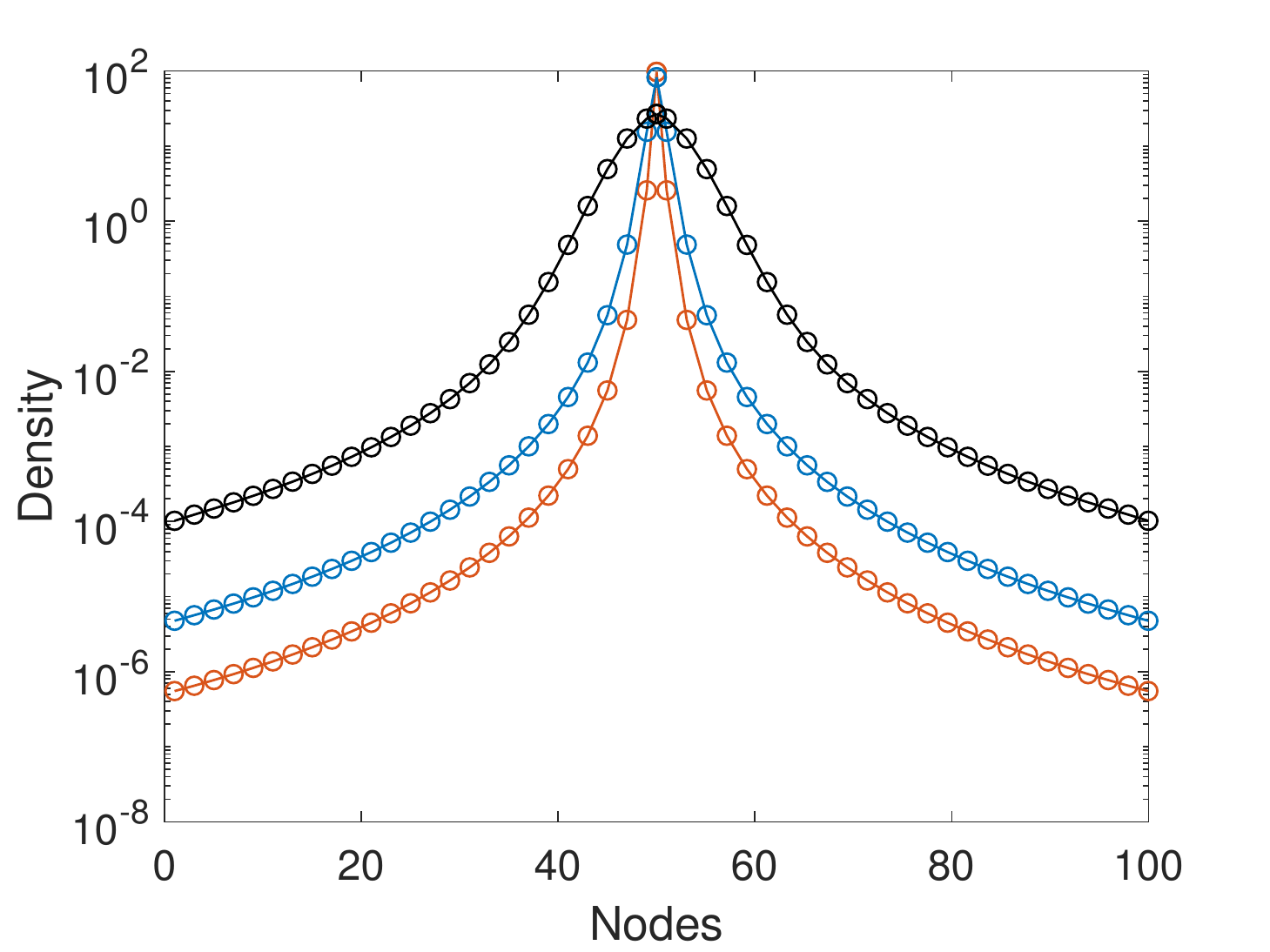}
\caption{Gaussian shape of density distribution for $s_{net}=4$ and $s_{nod}=0.8$ at $t=10$ (red), $t=100$ (blue) and $t=1000$ (black).}\label{fig: normal diffusion Mellin}
\end{figure}

In \cref{fig: normal diffusion Mellin} we exhibit the absence of superdiffusion for large Mellin exponent $s_{net}=4$ and $s_{nod}=0.8$: The density distribution recovers a Gaussian shape, characteristic of normal diffusion, as is clearly visible for longer times. Note that for the illustration of the Gaussian shape we have symmetrically reflected the network with respect to the $y$-axis.

Finally, \cref{fig: mellin densities} shows the equilibration of the density in node 1 with time. It confirms  the interplay of the nodal diffusion process and the size of the node,  as observed for the short ranged network diffusion above in \cref{fig: density different sizes}.

In closing, after all the analyses carried out in this section we can conclude that 
for the path graph $\mathcal{Q}$ and for $T_ {ij}$ given by the $d$-path Laplacian \eqref{kpathlap}, there exists $C>0$ such that $\int_{\Omega_j} u_j(t,x)\ dx \leq C e^{-C|j-1|}$ for $j \in \mathcal{Q}$. In particular, superdiffusion is not possible for the considered internal diffusion $\{H_k\}_k$.

 In the case of the Mellin-transformed $d$-Laplacian \eqref{kpathmel}, the numerical results indicate that superdiffusion in the network may persist even for nodes with distant sinks and sources. While in general the range of Mellin exponents $s_{net}$ which allow superdiffusion may shrink, for certain experiments in the linear network $V=\mathcal{Q}$ we numerically recover superdiffusion for $s_{net} \in (1,3)$, as in absence of internal structure \cite{estradalaa}.

\section{Real world metaplexes}
In this section we study the diffusion in two real-world
metaplexes of very distinctive type. The first, a landscape metaplex representing a fragmented forest from the south of Madagascar which shares some of the large-world properties of the linear metaplex in \Cref{sec: numerics}. A metaplex representing the  cortical region of a macaque, used as a second example, illustrates the dominance of the endo-dynamics in an ultra small-world network.

Here we 
focus on the diffusive processes on both systems. The analogies between
brain and ecological anomalous diffusion seem to be more than casual. Costa
et al. \cite{costa2016foraging} have coined the term ``foraging brain'' to refer to these
similarities, and the connection between animal foraging and cognitive
foraging has been considered by Hills \cite{Hills} on an evolutionary
basis.

In the following subsections we consider either small $\Omega_{s}$ or big $\Omega_{b}$
disks for all the nodes in the metaplex exactly as defined for the
toy model. The goal of this is to analyze whether the size of the
nodes have some influence on the global dynamics of the metaplex.
Thus, we can consider that $\Omega_{s}$ and $\Omega_{b}$ are
 average sizes of the regions in the corresponding metaplex,
such average is either relatively small or big.
 The coupling strength is given by $\alpha=10$ for the small nodes $\Omega_s$, and $\alpha=100$ for the big nodes $\Omega_b$. 
The nature of the coupling, as for the toy model in \Cref{sec: numerics}, is considered to be long ranged, proportional to $\textnormal{dist}(i,j)^{-s_{net}}$ according to the Mellin transformed $d$-path Laplacian \cref{kpathmel}.  The results for the short range coupling can be found in the Supplementary Material (Note 4).

For the landscape metaplex we discretize each node using a finite element method with 95 degrees of freedom, while for the metaplex of the cortical regions of a macaque we use a mesh of 347 degrees of freedom

The study of real-world metaplexes imposes some limits to the models
in use which were not needed for the idealistic toy model studied
before. The first is that the consideration of a unique source and sink
at the centre of each node seems unreasonable. For both cases, the
landscape and the visual cortex, we consider spatial regions which
can be connected to each other by disjoint sources and sinks as in \Cref{fig: density profiles b}.


\subsection{Landscape metaplex}\label{sec: landscape}
The first example considered represents a landscape in the south of Madagascar that has been fragmented into $183$ patches, corresponding to the nodes of a metaplex, as a result of agricultural activity \cite{bodin2006value}. These patches are connected by 529 narrow corridors that
creates the exo-structure of the metaplex.
 Such patches are vital for the survival of the ecosystem
because they are now the main habitat for the \textit{Lemur catta}, who plays a fundamental role as a major seeds propagator in this environment \cite{bodin2007network,estrada2008using}. 

Here we assume that all the patches have the same geometry.  The
network has relatively low edge density -- number of edges divided
by the maximum possible number of edges -- $\delta\approx0.032$ with about 45\% of nodes with degrees 2 and 3, and only two nodes
with degree 16. The
average separation between two patches in terms of the shortest path
distance is approximately $11.88$ with at least one pair of patches
separated at $d_{max}=32$. The average degree of the patches is approximately $5.78$
and the maximum number of nearest neighbors that a patch has is $16$.

\begin{figure}
\centering
\subfloat[]{\label{landscape metaplex}\includegraphics[scale=0.24]{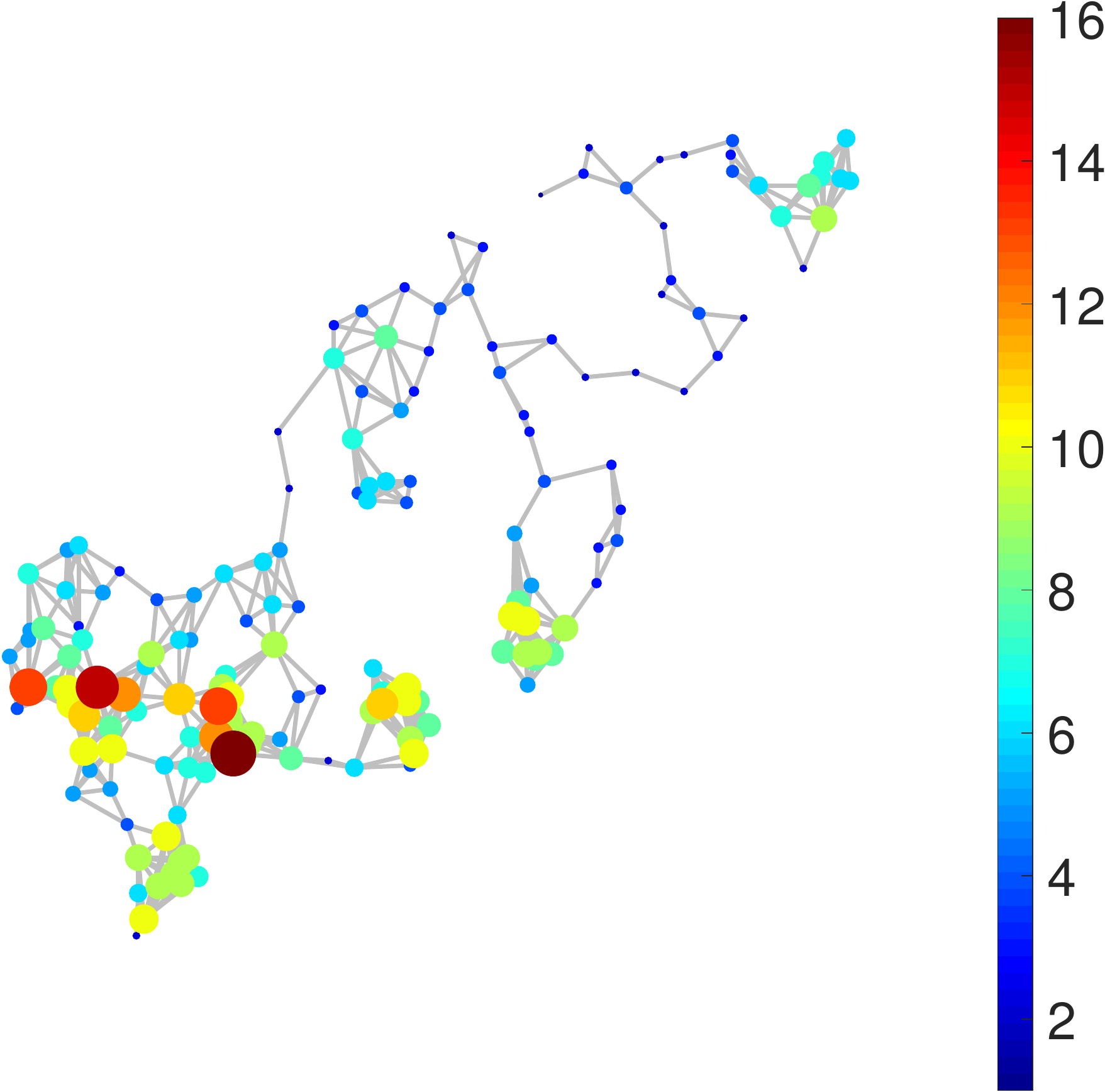}}
\hspace{1cm}\subfloat[]{\label{macaque metaplex}\includegraphics[scale=0.3]{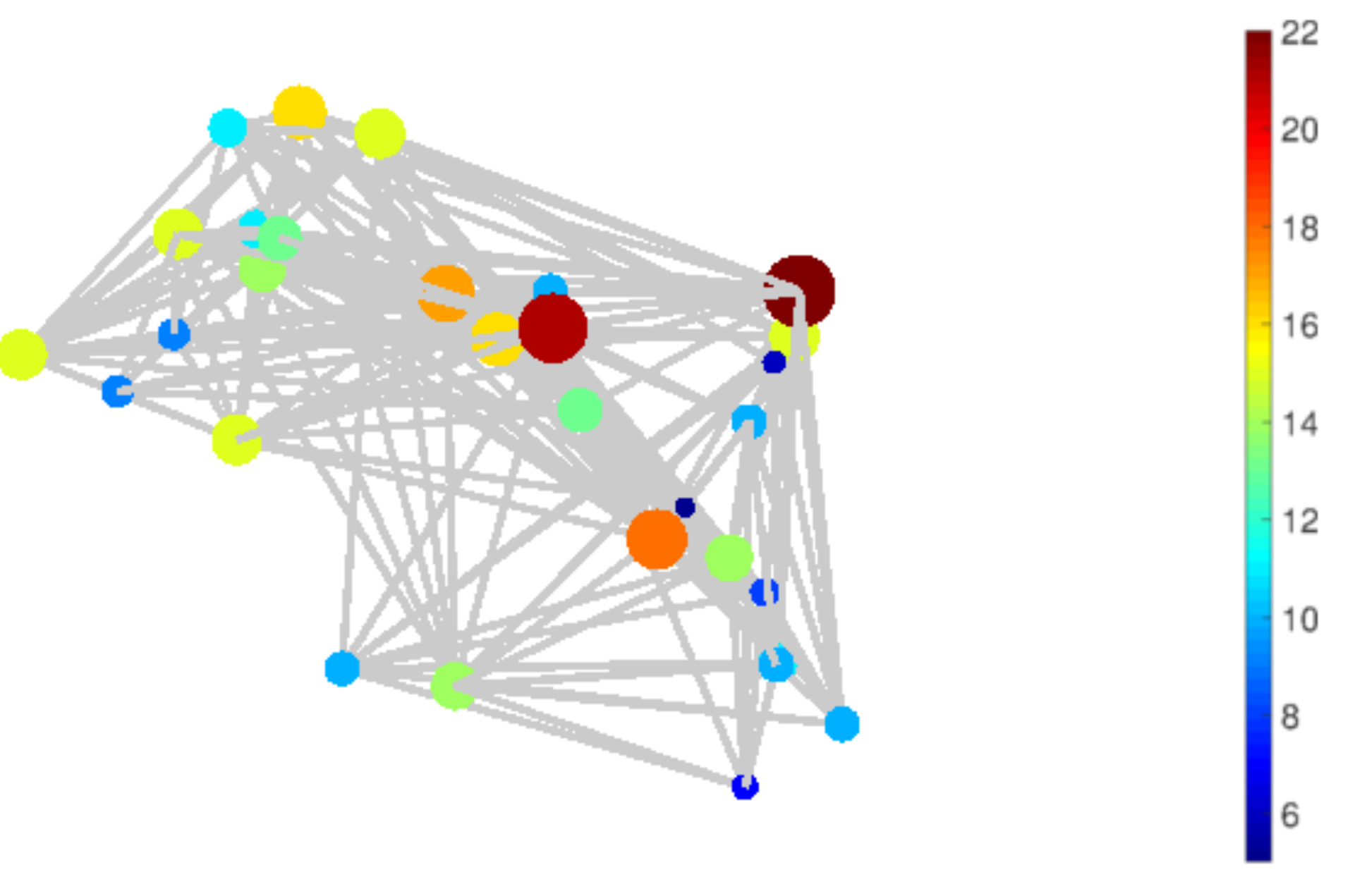}}
\caption{Illustration of the exo-skeleton of  a landscape (a) and a macaque metaplex (b) studied
here. The nodes are drawn with size and color proportional to
their degrees.}
\end{figure}

Also, we can observe that
most of the hubs -- high degree nodes -- are clumped together in a certain
region of the landscape (see \Cref{landscape metaplex}), while
the low degree nodes are relatively spread across the landscape. Consequently,
we have initialized the diffusive process by placing the initial condition
either into a randomly selected hub or into a randomly selected low-degree
node. 

The next experiment allows us to investigate whether
the degree of the node at which the diffusion starts makes an important
influence on the global metaplex diffusion. 
\begin{figure}[tbhp]
  \centering
  \subfloat[]{\label{fig: landscape one sink low}\includegraphics[scale=0.35]{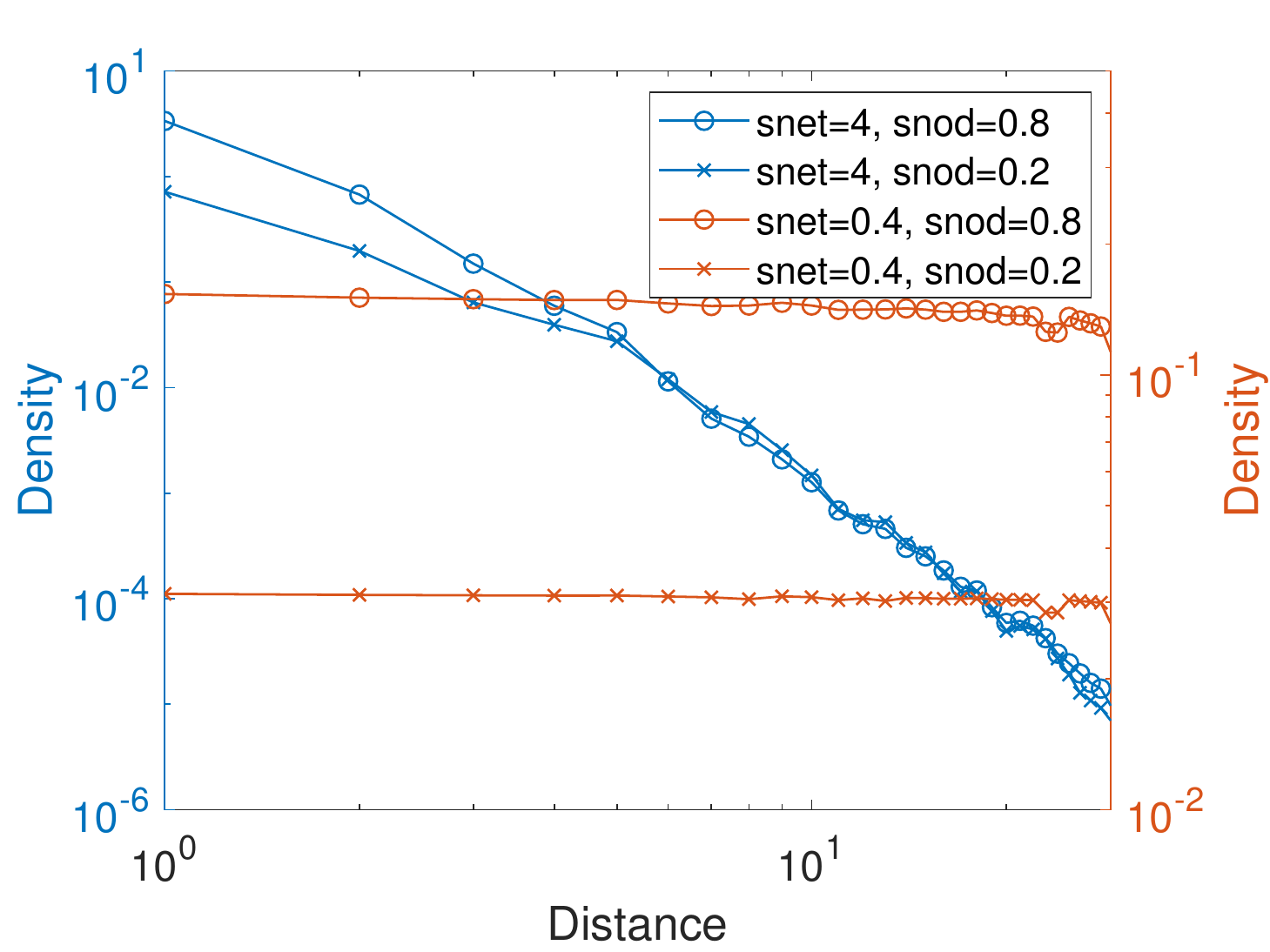}}
  \subfloat[]{\label{fig: landscape one sink high}\includegraphics[scale=0.35]{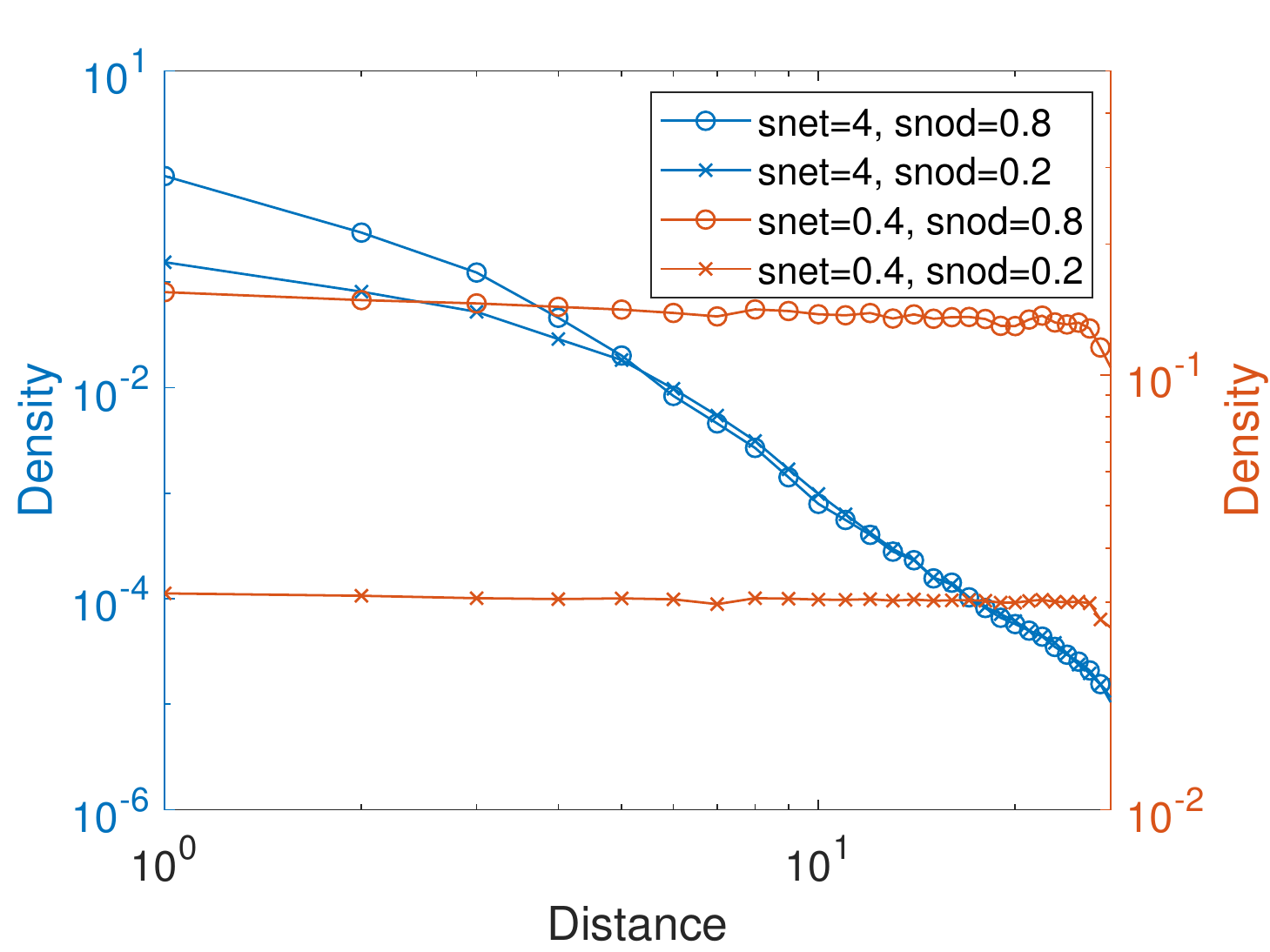}}  
  \caption{Density as a function of distance for disjoint  coupling point at $t=100$. (a) Low connectivity nodes. (b) High connectivity nodes.}\label{fig: landscape same sink}
\end{figure}

In \Cref{fig: landscape same sink} \textcolor{black}{ we plot the density of the diffusive particle
as a function of the shortest path distance from the initial condition when the diffusion starts at a low (}\Cref{fig: landscape one sink low}\textcolor{black}{) or at a high (}\Cref{fig: landscape one sink high}\textcolor{black}{)
degree node. Here
we averaged all the densities for nodes at exactly the same shortest
path distance from the initial condition where three different nodes with high (resp. low) connectivity where chosen to initialize the process. We observe that there are no significant differences between \Cref{fig: landscape one sink low} and \Cref{fig: landscape one sink high}. Only when $s_{net}=4$
there are some differences between the process initiated at low and
at high degree nodes. In this particular case, it seems that when
the process starts at a low degree node the density at nodes close
to the initial condition shares more concentration than those more
distant from it. {This localization effect can be the consequence not
of the degree of the node but of the relative geographic isolation
that such nodes can display in this landscape. That is, one low-degree
node is essentially surrounded by other low-degree nodes (see
}}\Cref{landscape metaplex}\textcolor{black}{). However, in all
the other cases it is important to remark that the profiles of the
density at different distances from the initial conditions are almost
exactly the same for processes starting at low or high degree nodes.
Consequently, we conclude that in this metaplex the place at which
the diffusion is initiated is not relevant for the evolution of the
global process. Therefore, in the rest of this section we discuss
the results for the process started from low-degree nodes.}

\begin{figure}[tbhp]
  \centering
  \subfloat[]{\label{fig: landscape diffsink low expon}\includegraphics[scale=0.34]{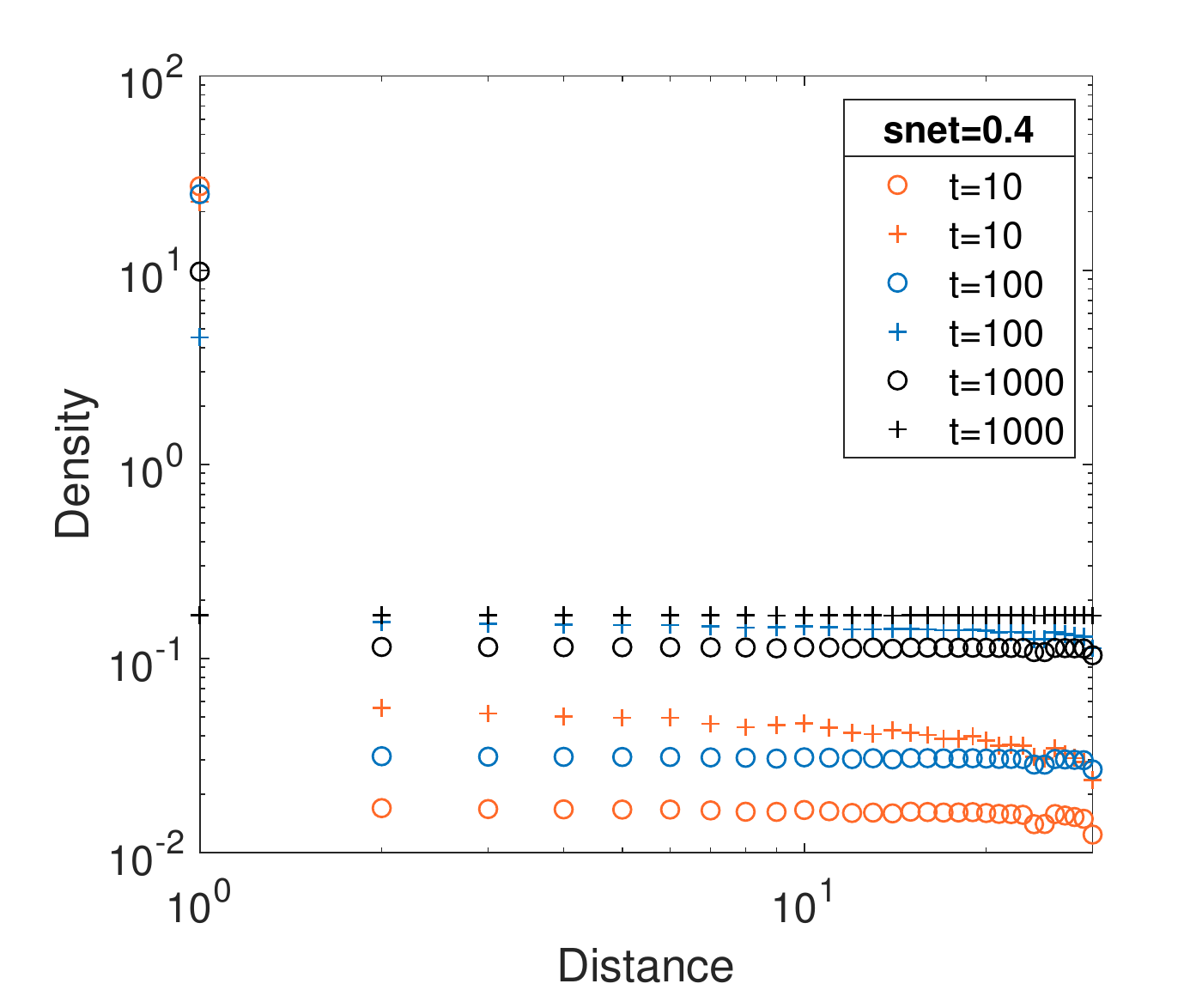}}
   \subfloat[]{\label{fig: landscape diffsink low mellin}\includegraphics[scale=0.34]{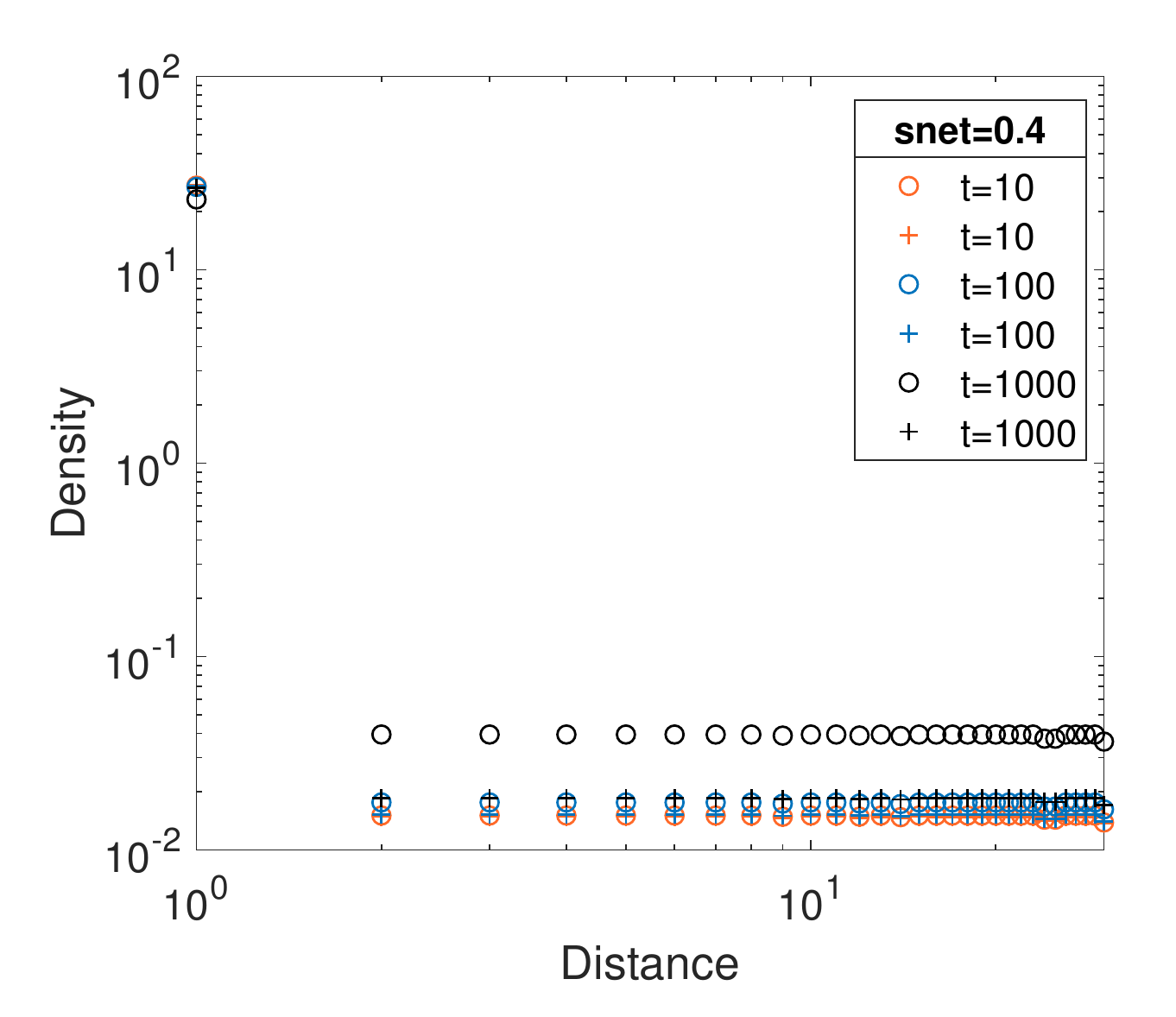}}
   
   \centering
    \subfloat[]{\label{fig: landscape diffsink expon high}\includegraphics[scale=0.34]{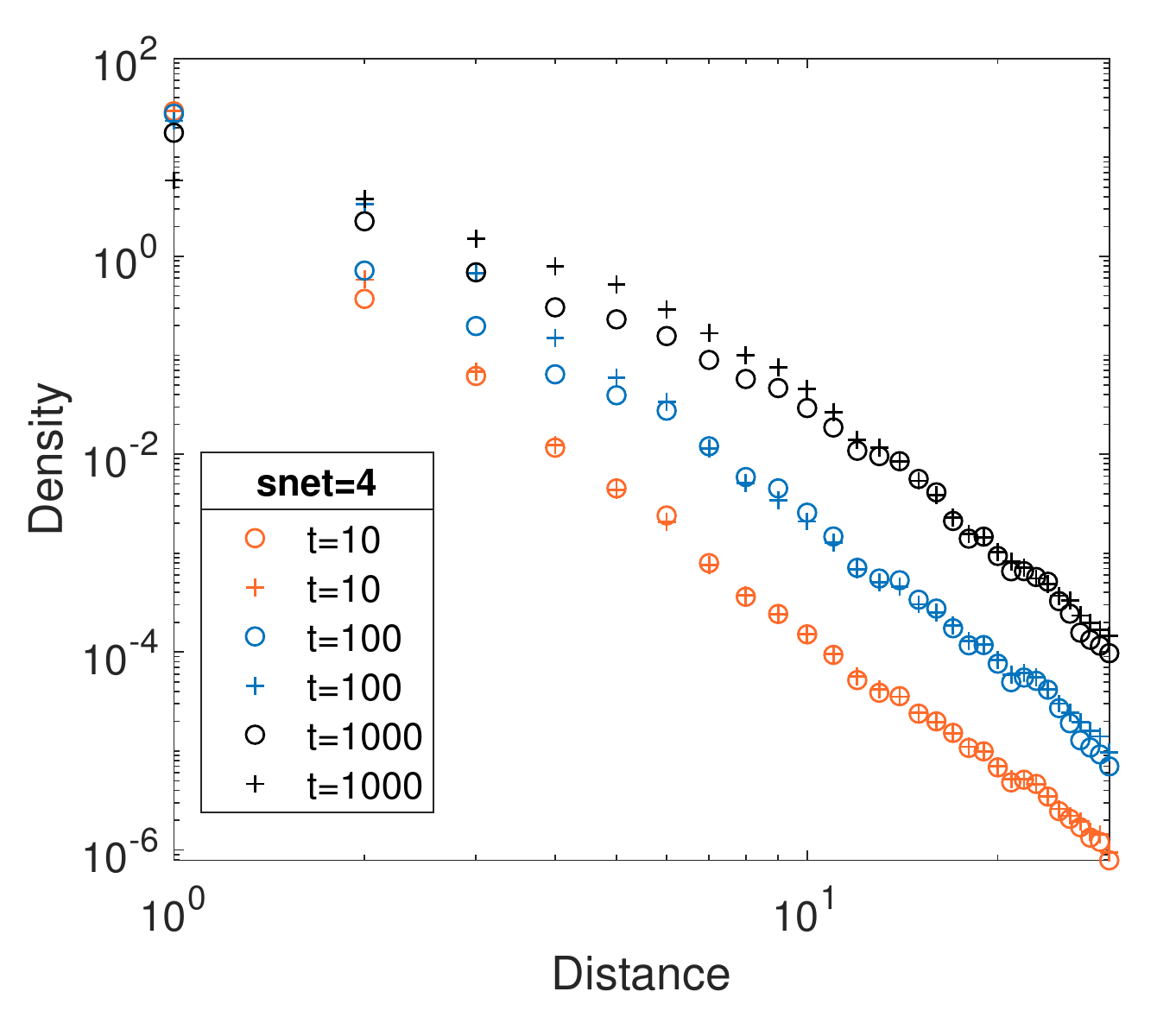}} 
      \subfloat[]{\label{fig: landscape bignode mellin low}\includegraphics[scale=0.34]{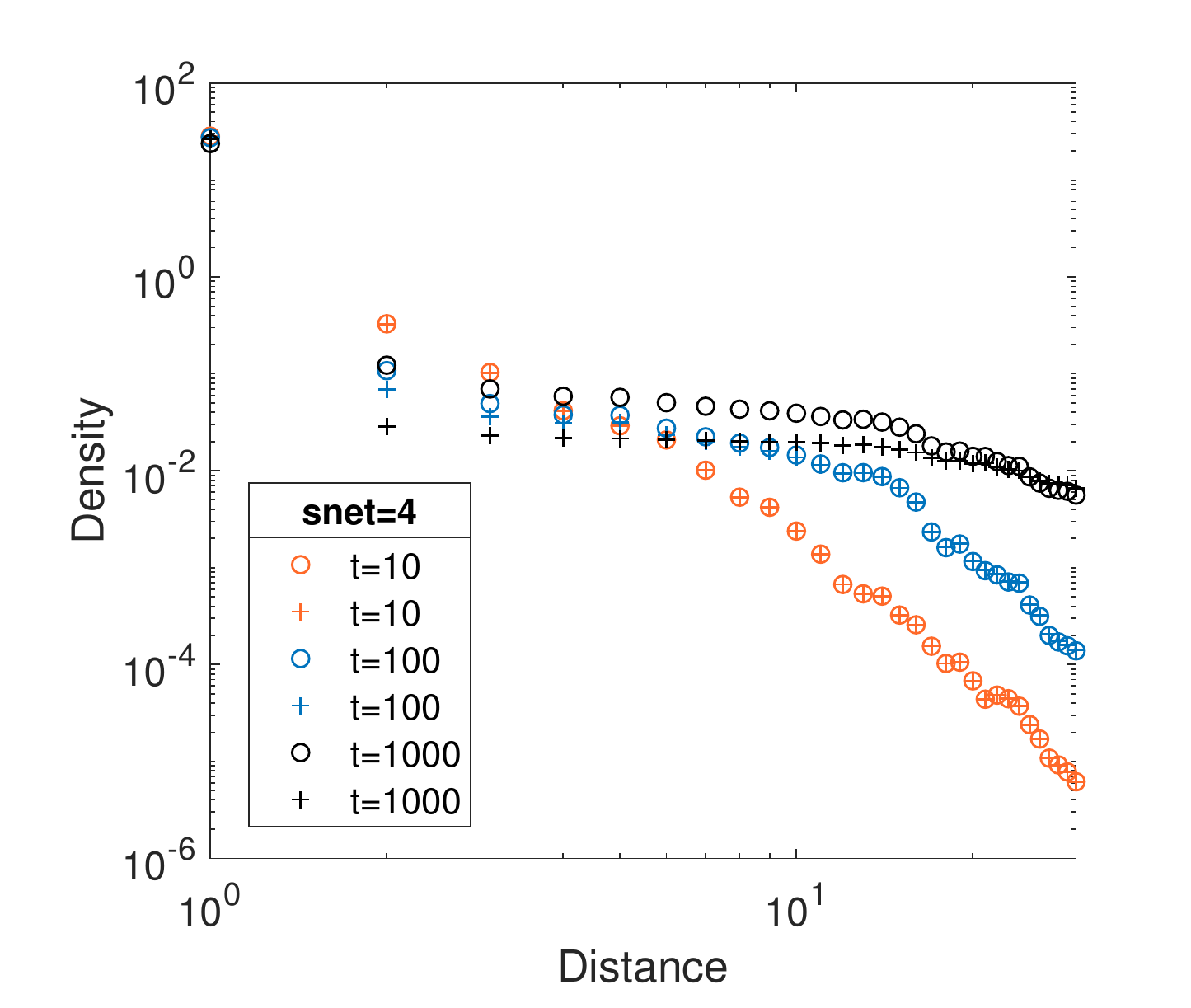}}
  
  \caption{Density as a function of distance for disjoint coupling points and $snod=0.8$ $(+)$, $snod=0.2$ $(\circ)$. (a) and (c) correspond to $\Omega_s$ and (b) and (d) correspond to $\Omega_b$.}
    \label{fig: landscape diffsink}
\end{figure}

We now explore the influence of the geometry of the nodes. 
In \Cref{fig: landscape diffsink} \textcolor{black}{we
plot the change of the density  for nodes
at different shortest path distances from the initial condition, for a small (\Cref{fig: landscape diffsink low expon}) and big node (\Cref{fig: landscape diffsink low mellin}) using
the same value of $s_{net}=0.4$. The first clear observation from these plots is the
following. When the nodes are small, superdiffusion inside the nodes slows down the rate to the equilibration of the diffusion
in the metaplex (notice that crosses are over the circles in } \Cref{fig: landscape diffsink low expon}\textcolor{black}{). However, when the nodes are big, the reverse
occurs, and relatively high values of $s_{nod}=0.8$ slow down the
equilibration of the diffusion in the metaplex (the circles are over
the crosses in } \Cref{fig: landscape diffsink low mellin}{.)
The
same results are reproduced when a much bigger value of $s_{net}=4$
is used, as can be seen in }\Cref{fig: landscape diffsink expon high,fig: landscape bignode mellin low}. 

\textcolor{black}{ The
reason for this apparently counterintuitive behavior has been previously
discussed for the toy model: in a small node it is counterproductive
to make long jumps inside the node. The reason is that the diffusive particle will rarely
find the sink to escape from the node. In a similar fashion, inside
a big node it is counterproductive to make small jumps as it will slow down the mobility of particles away from the source to reach the sink and leave
the node. We again conclude that the size of the nodes influences the global dynamics of the metaplex. Also importantly, the nature of the internal structure
can change significantly the rate of convergence of the process and
can make it faster or slower depending of the size of the nodes and
the nature of the diffusive process inside them.}

Let us now consider the combined influence of the endo- and exo-dynamics
on the global metaplex diffusion. We start here by considering the
time evolution of the diffusive particle across the metaplex after
initiating the process at a randomly selected node. 

\begin{figure}
\centering
\subfloat[]{\label{fig: L time evolution small node}\includegraphics[scale=0.35]{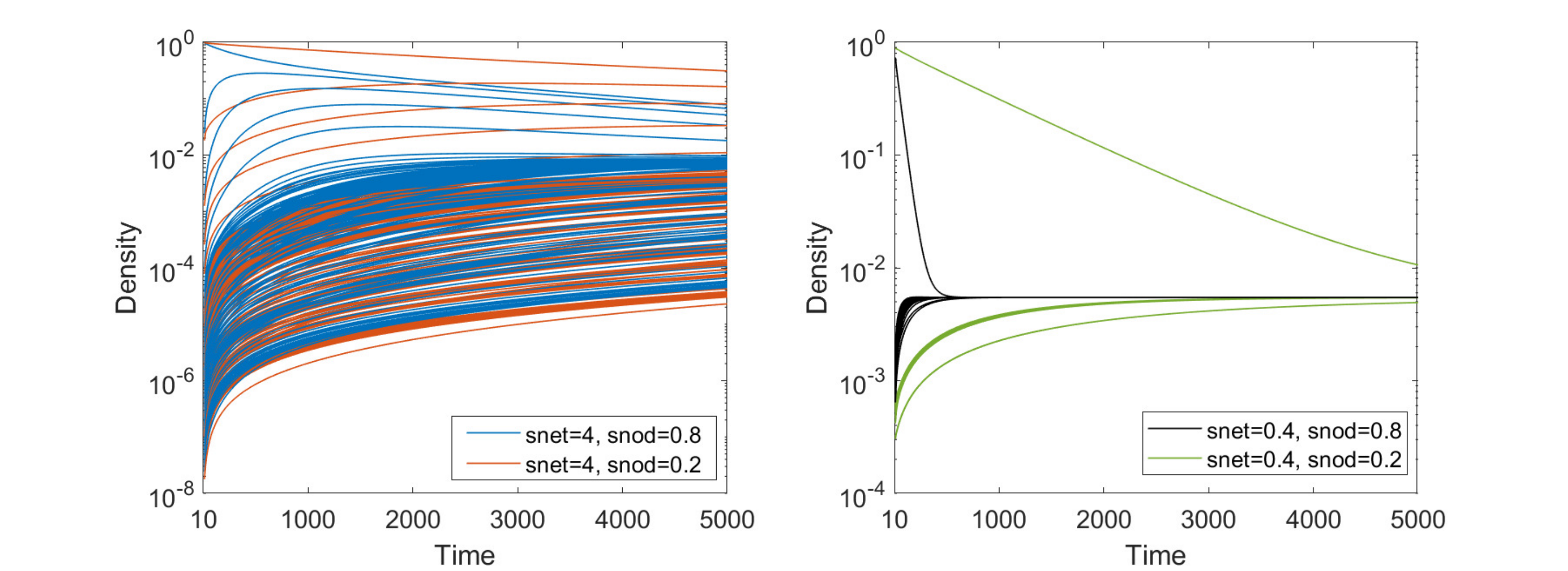}}

\subfloat[]{\label{fig: L time evolution big node}\includegraphics[scale=0.35]{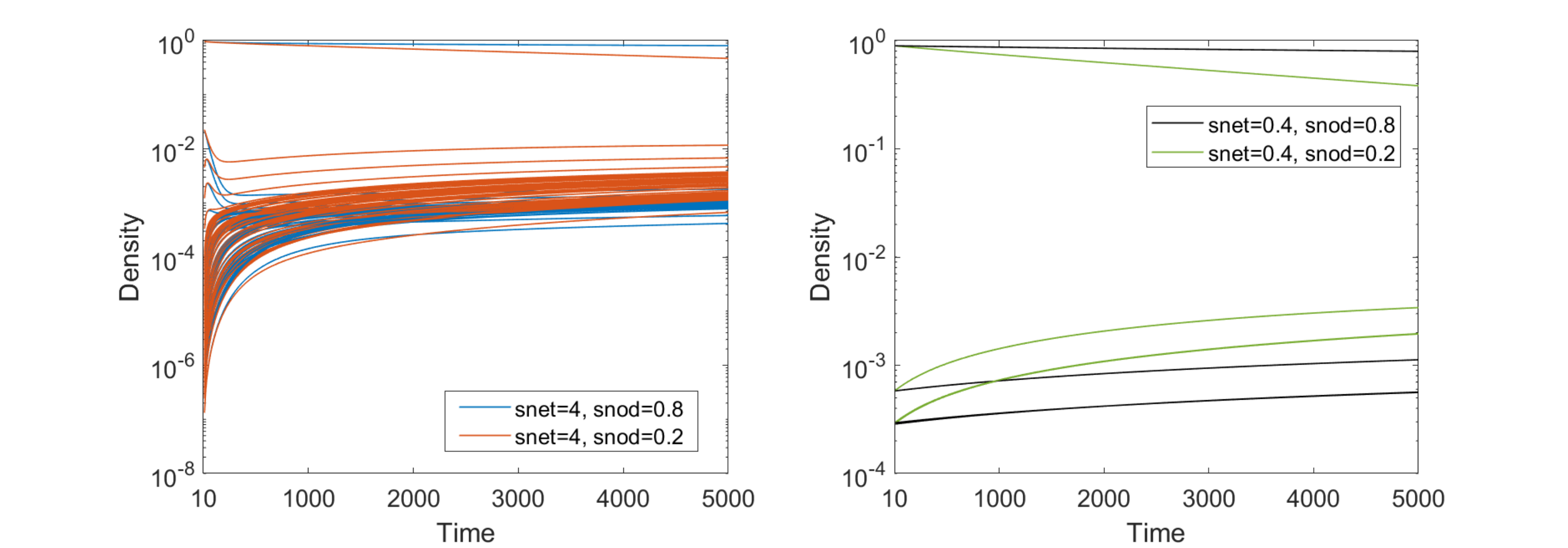}}

\caption{Time evolution of the diffusion dynamics for $\Omega_s$ (a) and $\Omega_b$ (b) with disjoint sink and sources inside the node of the landscape metaplex.}
\label{landscape evolution}
\end{figure}

In \Cref{fig: L time evolution small node} we illustrate the time evolution of the density (in logarithmic
scale) of the diffusive particle. 
{In the left panel of \Cref{fig: L time evolution small node}
we observe that the diffusion with $s_{nod}=0.8$ converges to the
steady state at an earlier time than that for $s_{nod}=0.2$ when
the same value of $s_{net}=4$ is used.  The same
is observed, even clearer, in the right panel where the process for
$s_{nod}=0.8$ and $s_{net}=0.4$ converges significantly faster
 to the steady state than that for
$s_{nod}=0.2$ and $s_{net}=0.4$ 
. See \Cref{tab:label} for the values of the standard deviation introduced in \Cref{sec: def}}.

\begin{table}[H]
\centering
\begin{tabular}{|c|c|c|c|c|}
 \hline
    & \multicolumn{2}{|c|}{$s_{net}=4$}  & \multicolumn{2}{|c|}{$s_{net}=0.4$}  \\
 \hline
 & $s_{nod}=0.2$& $s_{nod}=0.8$& $s_{nod}=0.2$& $s_{nod}=0.8$\\
 \hline
 $\Omega_s$   & 0.0265    &0.0088&   3.9$\cdot 10^{-4}$& 1.8$\cdot 10^{-11}$\\
 \hline
 $\Omega_b$&  0.034  & 0.059   &0.028& 0.059\\
 \hline
\end{tabular}
\label{tab:label}
\caption{Standard deviation at $t=5000$ for the landscape metaplex}
\end{table}


Making the nodes bigger shows the same behaviour as discussed above. The slower dynamics, compared to the small nodes, (see \Cref{fig: L time evolution big node})
is expected from the fact that now the diffusive particle has more
area to diffuse before finding the way out of the node. However, the
most interesting observation is that now, as we have previously pointed
out, the order of convergence rates is reversed. For the bigger node, the global diffusion equilibrates faster when we have superdiffusion inside the nodes than when $s_{nod}=0.8$ (compare values in \Cref{tab:label} for $\Omega_b$). Similar results are observed for the case $s_{net}=0.4$.

 From \Cref{fig: L time evolution small node} we observe that
\textcolor{black}{varying $s_{net}$  and $s_{nod}$ changes
the rate of convergence of the diffusion in the  metaplex.  So we conclude that, although the landscape metaplex has a large-world exo-skeleton,
it displays a trade-off between the exo- and endo-dynamics that determines
the global convergence of the diffusion in the metaplex. It is neither the exo-structure nor the endo-structure alone, but
a complex interrelation of both which controls the global dynamics
in the metaplex.}

\subsection{Visual cortex metaplex of macaque}

The second metaplex consists of 30 regions of the macaque visual cortex.
These cortex regions are functionally connected by 190 edges, which
makes the exo-structure of this system very dense, with density $\delta\approx0.437$.
As a consequence of this high edge density the network is a ``ultra-small-world''
with average shortest path distance between cortex regions of $\bar{d}\approx1.54$
and a maximum separation between two of these regions of only $d_{max}=3$
steps. This makes the exo-skeleton of this metaplex a ``ultra-small-world''. The average degree is approximately $12.67$ with a maximum
number of connections at a given node is $22$. 

This network displays an almost uniform
degree distribution, particularly for degrees between  $4$ and $16$. Therefore,
we just pick at random the nodes to initiate the diffusion on the
metaplex as most of the nodes are degree-equivalent.  
We then proceed with a similar analysis as for the case of the landscape
metaplex by studying the effects of the endo- and exo-dynamics on
the global diffusion.

\begin{figure}
\begin{centering}
\subfloat[]{\label{fig: M time evolution small node}\includegraphics[scale=0.35]{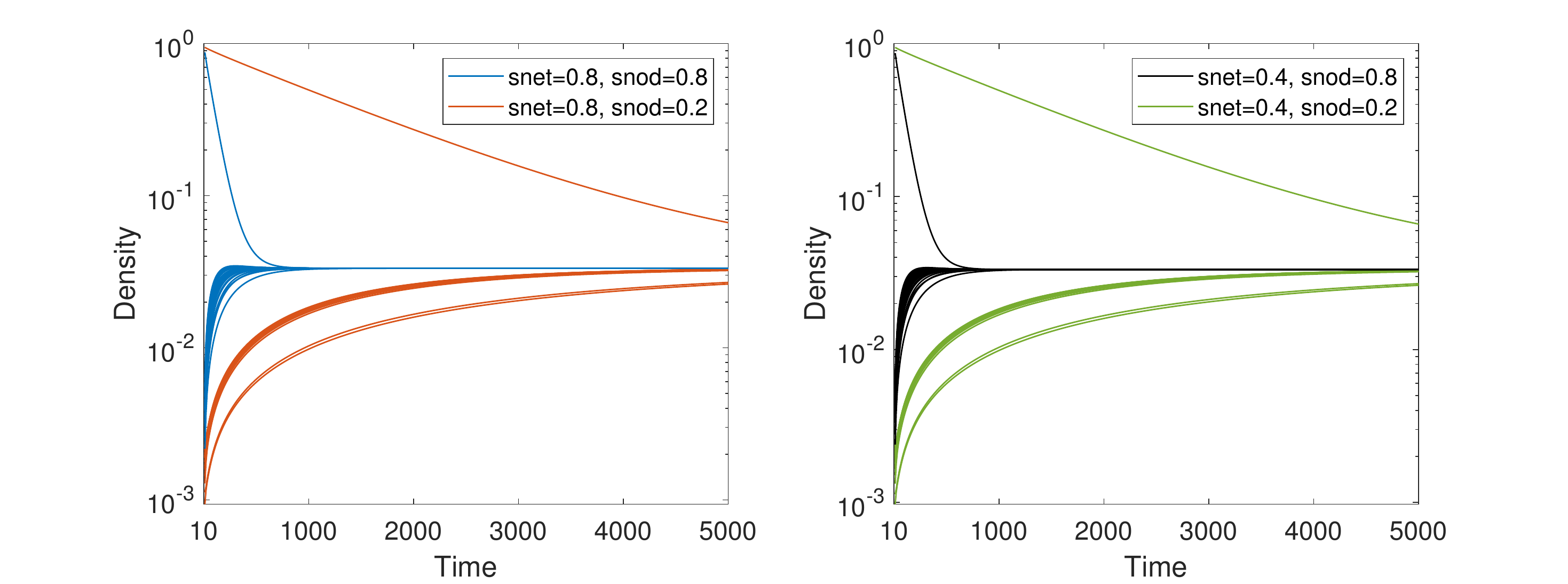}}

\subfloat[]{\label{fig: M time evolution big node}\includegraphics[scale=0.35]{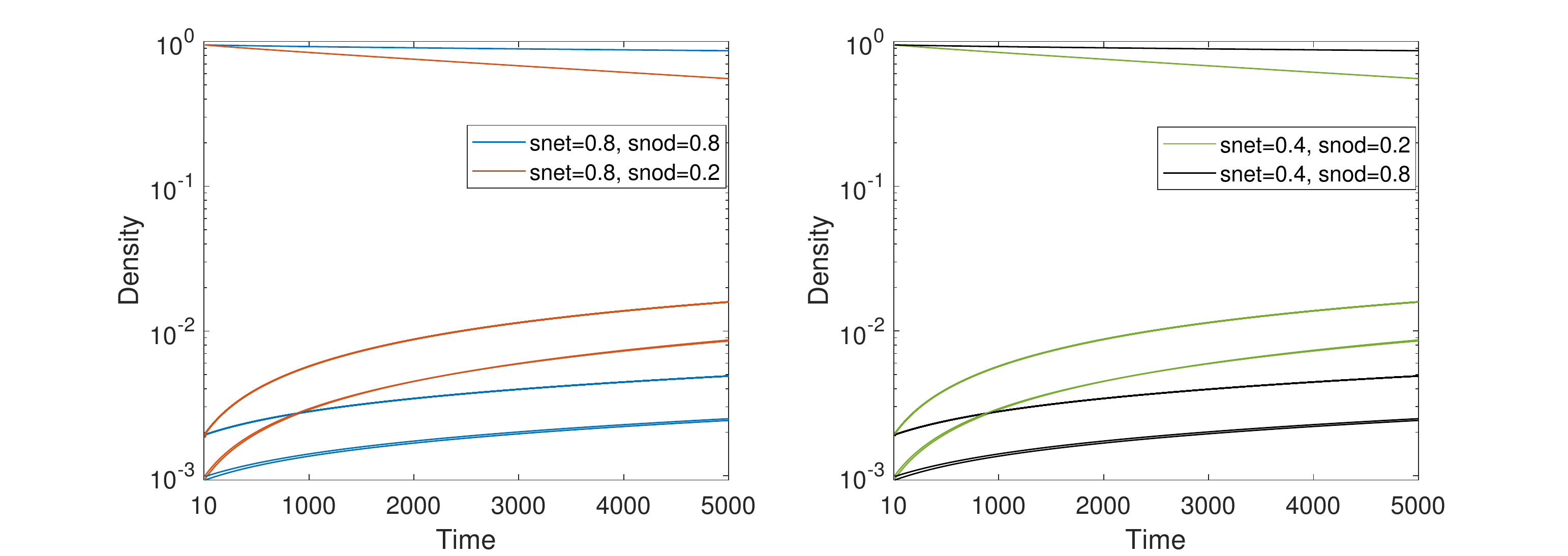}}
\par\end{centering}
\caption{Time evolution of the diffusion process on the macaque visual cortex
metaplex using different parameters for the endo- and exo-dynamics
(see text for details).}

\label{macaque evolution}
\end{figure}
\begin{table}[!ht]
\centering
\begin{tabular}{|c|c|c|c|c|}
 \hline
    & \multicolumn{2}{|c|}{$s_{net}=0.8$}  & \multicolumn{2}{|c|}{$s_{net}=0.4$}  \\
 \hline
 & $s_{nod}=0.2$& $s_{nod}=0.8$& $s_{nod}=0.2$& $s_{nod}=0.8$\\
 \hline
 $\Omega_s$   &   0.0064  & 1.378$\cdot 10^{-9}$ &   0.0064& 1.3283$\cdot 10^{-9}$\\
 \hline
 $\Omega_b$&  0.0983 & 0.1567   &0.0983& 0.1567\\
 \hline
\end{tabular}
\caption{Standard deviation at $t=5000$ for the macaque visual cortex metaplex}
\label{tab:label2}
\end{table}
When the nodes are small, \Cref{fig: M time evolution small node}, we observe the same effect as
in the previous cases, namely that larger values of \textcolor{black}{$s_{nod}$
favor the convergence of the dynamics to the steady state (see \Cref{tab:label2} for the standard deviation values). The results for the big nodes (} \Cref{fig: M time evolution big node})  are similar as for the toy model and landscape metaplex, i.e., superdiffusion inside the nodes{ favors
the convergence of the dynamics to the steady state.}

For the macaque visual cortex metaplex  the change of $s_{net}$
practically changes nothing in the global dynamics. For instance,
w{e  see from \Cref{fig: M time evolution small node} that keeping constant $s_{nod}=0.8$
and dropping $s_{net}$ from 0.8 (blue lines in the left panel) to
$s_{net}=0.4$ (black lines in the right panel) almost leaves unaffected
the time evolution of the diffusion. The standard deviations at $t=5000$
time units are in \Cref{tab:label2}.  We observe a similar behaviour for the case of big nodes (see }\Cref{fig: M time evolution big node}).

These results are in clear contrast with the ones previously discussed
for the landscape metaplex. In the previous case it was observed a
trade-off between the exo- and endo-dynamics to control the global
diffusion at the metaplexic scale.\textcolor{black}{{} Here, the exo-dynamics
does not play any significant role in the global dynamics of the metaplex.
The factor which determines such global dynamics is the endo-structure
of the nodes in the metaplex. The reasons for this extreme case of
dependence of the global dynamics almost exclusively is two-fold.
Firstly, the exo-skeleton of the macaque visual
cortex analyzed here is an ultra-small-world, where long range hops are not possible. The second is the almost-complete nature of
this exo-structure, i.e., its high density and uniformity of degree
distribution. It is intuitively clear that in a complete network there
are no topological effects that can affect the dynamics. In this case
only the endo-dynamics has a significant influence on the metaplexic
global dynamics.}

Most real-world metaplexes are expected to have exo-structures which
are not so dense and uniform as the one of the macaque visual cortex.
Although they certainly display small-world properties, they show
a larger variability of shortest path distances than the macaque cortex.
Thus, it is expected that they display certain
trade-off between the endo- and the exo-structures like the one observed
here for the landscape metaplex in determining the global dynamics
of the system.

\section{Summary}\label{sec: conclusion}

The concept of metaplex introduced in this work allows to study the
trade-off between the internal structure and dynamics of/in the nodes.
Here we provide the basic notions and motivations for the study of
metaplexes with continuous internal structure of the nodes and discrete
inter-node dynamics. Additionally, we provide theoretical and computational
support for a series of results related to the study of diffusive
dynamics on metaplexes. In particular, we present examples in which the endo structure of the metaplex determines almost uniquely the global dynamics: in the macaque visual cortex the exo-structure plays no fundamental role. On the other hand, in the linear metaplex chain there is a trade-off between the endo- and exo-dynamics to determine the global diffusion. This is reflected in the Madagascar metaplex landscape. 
We have also studied here the effect of the geometry, such as the size of nodes, location of sink and sources, and the nature and strength of the coupling between nodes.


Our numerical results show that superdiffusion due to long range hopping {in the linear network $\mathcal{Q}=\{\dots, -2,-1,0,1,2,\dots\}$, as in \cite{estradalaa}, survives irrespective of the internal structure of the nodes  (\Cref{fig: mellin superdiffusion}).} The parameter range of Mellin exponents for which superdiffusion is observed, $s_{net} \in (1,3)$, is the same as for ordinary networks of point nodes without internal structure.
{The conclusion extends to certain networks from applications, of a large diameter, here studied for the landscape network of habitats of \emph{L.~catta} (\Cref{fig: landscape diffsink}). The combine influence of the endo- and exo-structure determined the global diffusion in such networks.}

{The endo-structure, on the other hand, dominates at shorter distances and therefore becomes crucial for diffusion in small-world metaplexes. This was illustrated  for the cortical metaplex of the macaque,  \Cref{macaque evolution}.} The effect of the internal dynamics in the whole metaplex strongly depends on the geometry of the nodes and the nature of the coupling. When sinks and sources overlap, internal superdiffusion may {\emph{slow down}} the metaplex dynamics, and normal diffusion is faster on small scales. When sinks and sources are in separate, distant locations, superdiffusion in the nodes allows particles to explore the entire metaplex much faster than classical diffusion.  While it accelerates diffusion, the internal superdiffusion cannot, by itself, induce superdiffusion in the metaplex.  

Our results can be understood from the local description we provide of the distribution of particles inside each node. The combination of analytical methods for the PDE description in the node, and network methods for their interconnection in the exo-skeleton gives a new perspective not only on classical complex network descriptions, but also for the study of PDEs describing physical systems which can be either split into continuous regions interconnected in a discrete way or involve a network of internal degrees of freedom. This illustrates how the study of metaplexes will draw tools from a wide range of areas, such as the geometric analysis of coupled PDEs and interface problems, spectral theory of operator matrices, high-dimensional stochastic processes etc.  Conversely, it suggests the relevance of network theory for old problems in these fields, such as the problem of finding effective descriptions of interacting many-particle systems with high-dimensional internal degrees of freedom. 

From a computational point of view, numerical experiments for large metaplexes become a challenge, due to the new internal degrees of freedom in each node. For applications to real-world networks, future work should explore methods which represent this internal dynamics efficiently. Model order reduction or generalized finite element methods \cite{pufem} are examples which have been used in related settings to achieve a reasonable accuracy already for small degrees of freedom.


\appendix

\bibliographystyle{plain}
\bibliography{networks}

\end{document}